\documentclass[acmtog,anonymous]{acmart}
\acmSubmissionID{2152}

\acmJournal{TOG}

\begin{document}
	
	\title{Plug-and-Play PDE Optimization for 3D Gaussian Splatting: Toward High-Quality Rendering and Reconstruction}
	
	\author{Yifan Mo}
	\email{moyf@mail.ustc.edu.cn}
	\affiliation{%
		\institution{USTC}
		\city{Hefei}
		\country{CN}
	}
	
	\author{Youcheng Cai}
	\affiliation{%
		\institution{USTC}
		\city{Hefei}
		\country{CN}}
	\email{caiyoucheng@ustc.edu.cn}
	
	\author{Ligang Liu}
	\affiliation{%
		\institution{USTC}
		\city{Hefei}
		\country{CN}}
	\email{lgliu@ustc.edu.cn}
	
	\renewcommand{\shortauthors}{Mo et al.}
	
	
	\begin{abstract}
		3D Gaussian Splatting (3DGS) has revolutionized radiance field reconstruction by achieving high-quality novel view synthesis with fast rendering speed, introducing 3D Gaussian primitives to represent the scene. However, 3DGS encounters \textbf{blurring} and \textbf{floaters} when applied to complex scenes, caused by the reconstruction of \textbf{redundant} and \textbf{ambiguous} geometric structures. We attribute this issue to the unstable optimization of the Gaussians. To address this limitation, we present a plug-and-play PDE-based optimization method that overcomes the optimization constraints of 3DGS-based approaches in various tasks, such as novel view synthesis and surface reconstruction. Firstly, we theoretically derive that the 3DGS optimization procedure can be modeled as a PDE, and introduce a viscous term to ensure stable optimization. Secondly, we use the Material Point Method (MPM) to obtain a stable numerical solution of the PDE, which enhances both global and local constraints. Additionally, an effective Gaussian densification strategy and particle constraints are introduced to ensure fine-grained details. Extensive qualitative and quantitative experiments confirm that our method achieves state-of-the-art rendering and reconstruction quality.
		
	\end{abstract}

	\begin{teaserfigure}
		\includegraphics[width=\linewidth]{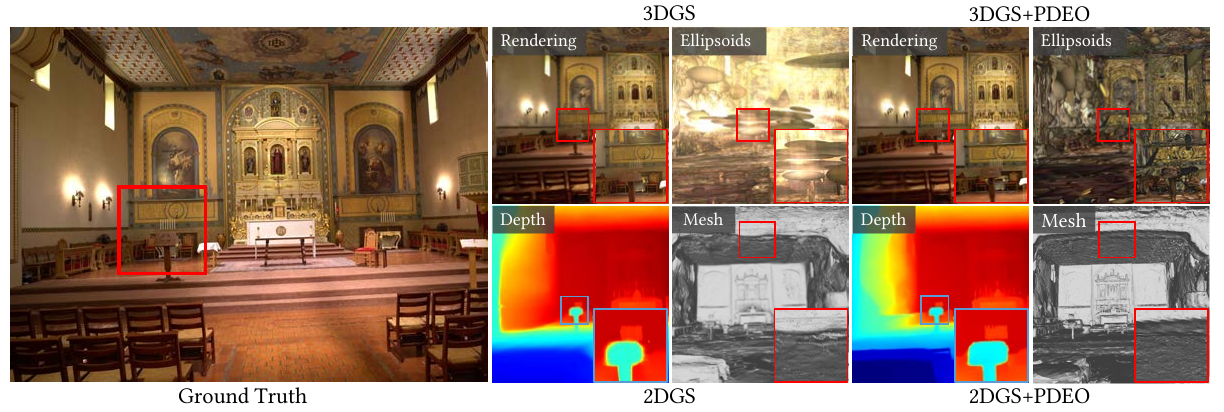}
		\caption{We present PDEO, a novel, plug-and-play optimization framework designed to enable stable optimization of 3D Gaussians and enhance existing 3DGS-based approaches for tasks such as novel view synthesis and surface reconstruction. Our method achieves high-quality results in both rendering and reconstruction. More results are provided in the accompanying video.}
		\Description{teaser}
		\label{fig1}
	\end{teaserfigure}

	
	
	\received{20 February 2007}
	\received[revised]{12 March 2009}
	\received[accepted]{5 June 2009}
	
	\maketitle

	\section{Introduction}
	The reconstruction of 3D scenes from multi-view images is a classic problem in computer vision and computer graphics. Recent advances in Neural Radiance Fields (NeRF) \cite{RN1} have revolutionized this task by introducing implicit neural representations, achieving state-of-the-art results. A notable follow-up is 3D Gaussian Splatting (3DGS) \cite{RN5}, which has gained increasing attention due to its high-quality, real-time rendering performance, attributed to its explicit point-based representation and efficient splatting process. 
	
\begin{figure}[t]
		\centering
		\includegraphics[width=\linewidth]{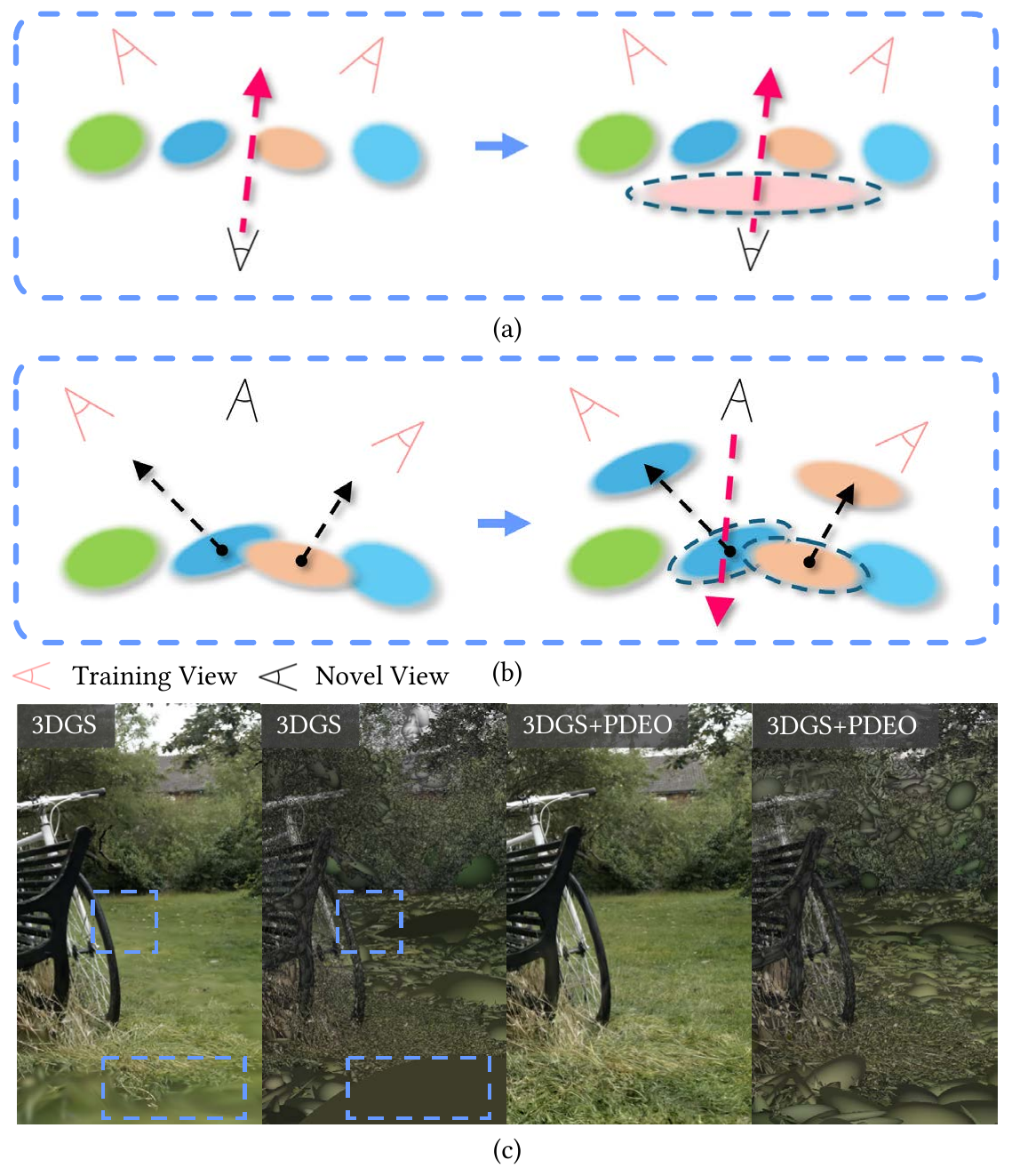}
		\caption{Optimization of 3D Gaussians. (a) The redundantly of large 3D Gaussians. (b) The ambiguity of small 3D Gaussians. (c) Visualization results.}
		\Description{}
		\label{fig2}
	\end{figure}

	
	When applied to complex scenes, 3DGS encounters \textbf{blurring} and \textbf{floaters}, as validated in Figure \ref{fig1}, resulting in degraded \textbf{rendering} and \textbf{reconstruction} quality. As shown in Figure \ref{fig2}(a), 3DGS tends to employ large Gaussians to fill voids in the scene, which struggle to accurately represent high-frequency details, resulting in over-reconstruction and visible blurring. Although small Gaussians are more effective at capturing high-frequency scene details, they tend to introduce numerous floaters, as demonstrated in Fig. \ref{fig2}(b). Regions with limited scene coverage tend to produce floaters in novel views, as the Gaussians are optimized to align with the training views. Existing works \cite{RN30,RN31} propose dividing large Gaussians into a greater number of smaller Gaussians using effective densification criteria. These methods employ an adaptive approach by fitting the scene with an excessive number of Gaussians, which is neither storage-efficient nor effective for rendering.
	
	Through intensive study, we have identified the reason why small Gaussians are prone to unstable optimization. According to the gradient computation, the magnitude of the positional gradient is significantly higher than that of the other attribute gradients when the Gaussian scale is small. Consequently, 3DGS tends to move these small Gaussians to fit the scene, thereby hindering the optimization of other Gaussian attributes during the optimization process. This abrupt positional change results in redundant and ambiguous geometric structures. To ensure stable gradient optimization, existing gradient optimization methods typically emphasize gradient clipping \cite{gc1,gc2}, normalization \cite{bn1,bn2}, and weight decay \cite{wd1,wd2}. However, these methods are heuristic in nature and inevitably lead to information loss. 
	
	In this paper, we aim to enable 3DGS to bypass its original optimization weaknesses and achieve more efficient and stable optimization. Building on the above observation, we propose the following insights: (1) The 3DGS optimization procedure can be modeled as the discretization of a partial differential equation (PDE). In this formulation, the attributes of the 3DGS are treated as functions of time. (2) Inspired by fluid simulation \cite{RN14}, we introduce a viscous term into the PDE to suppress abrupt motion changes and achieve stable optimization. The viscous term, which constrains particles through the local average velocity, effectively prevents abrupt changes in the motion of particles.

	We propose a novel, plug-and-play optimization framework based on PDEs, termed PDEO, that enhances existing 3DGS-based approaches for tasks such as novel view synthesis and surface reconstruction. \textbf{The goal is to adapt large Gaussians into smaller ones to better capture high-frequency details, and to enable stable optimization of small Gaussians for improved rendering and reconstruction quality.} Firstly, we theoretically derive that the 3DGS optimization procedure can be modeled as a PDE, and introduce a viscous term to ensure stable optimization. Secondly, we employ the Material Point Method (MPM) \cite{RN16} to solve the PDE, thereby enforcing both global and local constraints for optimization. Finally, we propose explicit particle constraints to enforce small-scale, high-confidence Gaussians in accordance with the particle hypothesis and an effective Gaussian densification strategy to to ensure fine-grained details. Extensive experiments demonstrate that our PDEO improves upon state-of-the-art methods as a plug-and-play optimizer, consistently enhancing performance in both novel view synthesis and surface reconstruction.
	
	In summary, the main contributions are provided as follows:
	
	\begin{itemize}
		\item We propose a novel, plug-and-play optimization framework based on PDEs, which enhances existing 3DGS-based approaches in novel view synthesis and surface reconstruction.
		\item We formulate the 3DGS optimization procedure as a PDE and introduce a viscous term to ensure stable optimization of Gaussians.
	\end{itemize}
	
	\section{Related Work}
	
	\subsection{Novel View Synthesis}
	The recent success of Neural Radiance Fields (NeRF) \cite{RN1} introduces an implicit scene representation that achieves high rendering quality in novel view synthesis. Subsequent methods \cite{RN20,RN21,RN22,RN23} are proposed to improve the original NeRF. For instance, Mip-NeRF \cite{RN20} proposes a new feature representation of the integrated positional encoding to improve the rendering quality. Later, MipNeRF360 \cite{RN20} extends this method to unbounded scenes by using a non-linear scene parameterization. Another line of work \cite{RN24,RN25,RN26} focuses on improving the efficiency of NeRF, which proposes to accelerate training and rendering by introducing volumetric features \cite{RN27,RN28} or sparse hash-based grids \cite{RN29}.
	
	3D Gaussian Splatting (3DGS) \cite{RN5} improves training and rendering speed by introducing anisotropic 3D Gaussians and efficient splatting, which supports forward rasterization and avoids the shortcomings of expensive sampling and queries. Some subsequent works on 3DGS further enhance performance for novel-view synthesis. For example, AbsGS \cite{RN30} and Fregs \cite{RN31} enhance the densification strategy of 3DGS to achieve more accurate density adjustments. GES \cite{RN32}, DisC-GS \cite{RN33} and 3D-HGS \cite{RN34} refine the basis function representation of 3D Gaussians to provide a more precise and detailed representation. Recently, 3DGS$^2$ \cite{lan20253dgs} proposes a second-order convergent training algorithm for 3DGS, which achieves a tenfold increase in training speed.

	
	\subsection{Neural Surface Reconstruction}
	
	Due to the absence of surface constraints, NeRF cannot extract high-quality surfaces. NeuS \cite{RN46} introduces a Signed Distance Field (SDF) to represent the geometric surfaces of the scene and improves the rendering formulations to achieve more accurate results. Neuralangelo \cite{RN47} introduces hash encoding into the SDF to enable detailed large-scale scene reconstruction. Binary Opacity Grids \cite{RN49} employ a discrete opacity grid to represent the scene, allowing for a more accurate representation. However, these methods continue to demand substantial training time owing to the high computational cost of volume rendering. 
	
	Recently, various studies \cite{RN15,RN16,RN56,RN57} have extended 3DGS to surface reconstruction. For instance, SuGaR \cite{RN59} introduces a regularization term that encourages the 3D Gaussians to align with the surface, facilitating more effective mesh extraction. GOF \cite{RN58} proposes a ray-tracing-based volume rendering approach to enable direct extraction of geometry from unbounded scenes. 2DGS \cite{RN56} and RaDeGS \cite{RN15} approximate surfaces with Gaussians by imposing shape constraints and incorporating depth information. Different from these methods that introduce explicit geometric constraints, our method uses a PDE-based optimization strategy, achieving more stable optimization and effectively eliminating redundant and ambiguous geometric structures.

	\subsection{Gradient Optimization}
	During the optimization process, it is not uncommon for one gradient to be significantly larger than the others, a phenomenon known as gradient explosion, which is a widely known issue in optimization. Gradient clipping \cite{gc1} is a widely used technique that constrains the gradient by applying an upper limit to the gradient magnitude. Subsequent works \cite{gc2,gc3} have built upon this approach by introducing more adaptive truncation methods. Batch normalization (BN) \cite{bn1,bn2} constrains the gradient by normalizing the attributes through a transformation, thereby facilitating a more stable optimization process. Weight decay \cite{wd1,wd2} modifies the loss function by adding a stabilizing term to constrain the gradient. Although these methods impose reasonable constraints on the gradient, they inevitably result in information loss in the original gradient. In contrast, our approach modifies the governing PDE to stabilize the gradient while preserving the integrity of the original gradient information. 
	
	\subsection{Material Point Method}
	Simulating natural phenomena for virtual worlds is a crucial application that remains extremely challenging. The Material Point Method (MPM) \cite{RN64} has been demonstrated to be an effective hybrid particle/grid method for simulating various solid/fluid materials in the solution of a partial differential equation (PDE), emerging as a generalization of the Particle-in-Cell (PIC) and Fluid Implicit Particle (FLIP) methods \cite{RN16}. MPM methods combine Lagrangian material particles \cite{RN65, RN66} with Eulerian Cartesian grids \cite{RN67,RN68}, which discretizes the initial PDE problem using material particles. For example, Stomakhin et al. \cite{RN69} employ the MPM to simulate snow, producing convincing results. Yue et al. \cite{RN70} demonstrate that MPM is also suitable for simulating complex fluids, such as foams. In this work, we propose PDE-GS, which models 3DGS optimization as a PDE, thereby introducing the MPM to solve the 3DGS optimization and achieve more stable and efficient optimization.

	\section{Preliminary and Motivation}
	\subsection{Preliminary}
	\subsubsection{3D Gaussian Splatting.} 3DGS \cite{RN5} employs a set of learnable 3D Gaussians that encapsulate surrounding information to represent the scene explicitly. Each 3D Gaussians $g_i$ is parameterized by learnable attributes of center position $\boldsymbol{\mu}_i$, opacity $\hat{o}_i$, color $\hat{\mathbf{c}}_i$ and a covariance matrix $\boldsymbol{\Sigma}$,
	
	\begin{equation}\label{eq:1}
		g_i(\boldsymbol{\mu})=\exp({-\frac{1}{2}(\boldsymbol{\mu}-\boldsymbol{\mu}_i)^T\boldsymbol{\Sigma}^{-1}(\boldsymbol{\mu}-\boldsymbol{\mu}_i)})
	\end{equation}
	where the covariance matrix $\boldsymbol{\Sigma}$ is denoted by the rotation matrix $\boldsymbol{R}$ and the scaling matrix $\boldsymbol{S}$ as $\boldsymbol{\Sigma}=\boldsymbol{RS}\boldsymbol{S}^T\boldsymbol{R}^T$.
	
	To render an image, the 3D Gaussians are projected onto the image plane and converted into 2D Gaussians through the splatting operation \cite{RN71}. Subsequently, the color $C$ of a pixel is computed by combining $N$ ordered Gaussians using $\alpha$-blending,
	\begin{equation}\label{eq:2}
		C=\sum_{i{\in}N}\hat{\mathbf{c}}_i\alpha_{i}{\prod_{j=1}^{i-1}(1-\alpha_j)}
	\end{equation}
	where $\alpha_j$ is computed by the 2D Gaussian multiplied with the opacity $\hat{o}_i$.
	
	
	\subsubsection{Material Point Method.} MPM \cite{RN16} is a discrete method for solving PDE, widely used in solid and fluid simulation \cite{RN14}. MPM combines the two perspectives of the system:  the Lagrangian description and the Eulerian description. In the Lagrangian description, the system is regarded as a discrete phase comprising numerous independent particles, each endowed with its own attributes. In contrast, the Eulerian description treats the system as a continuum phase, which enables a global description of the particle motion. 
	
	Specifically, the motion equation of particles evolves over time $t$ as: $f(\boldsymbol{v},\boldsymbol{x})=\frac{\partial\boldsymbol{v}}{\partial{t}}$, where $\boldsymbol{v}$ is velocity and $\boldsymbol{x}$ is position. Then, MPM is used to discrete the function as: $	f(\boldsymbol{v},\boldsymbol{x})=\boldsymbol{v}^{t+1}-\boldsymbol{v}^{t}$.

	\subsection{Gradient Analysis}
	3DGS employs gradient descent for scene optimization, a process that is essential for achieving high-quality scene representation. Each Gaussian $g_i$ is associated with a set of trainable attributes $\Gamma_i^t=\{\boldsymbol{\mu}_i, \mathbf{c}_i, o_i, \mathbf{s}_i, \mathbf{q}_i\}$, where $\boldsymbol{\mu}_i$ denotes the center position, $\mathbf{c}_i$ represents the spherical harmonic coefficients, $o_i$ is the opacity attribute, $\mathbf{s}_i$ refers to the scale attributes, and $\mathbf{q}_i$ is the quaternion representing the rotation attributes. Here, $\hat{o}_i = \operatorname{Sig}(o_i)$ denotes the opacity, where $\operatorname{Sig}(\cdot)$ represents the sigmoid function. $\hat{\mathbf{s}}_i = \exp(\mathbf{s}_i)$ denotes the scaling vector. During optimization, the update of each attribute is given by, 
	\begin{equation}\label{eq:3}
		{\triangle}\gamma_i=\sigma\frac{\partial{L}}{\partial{\gamma_i}}, \gamma_i\in\Gamma_i
	\end{equation}
	where $\sigma$ denotes the learning rate and ${L}$ represents the loss function. For simplicity, we consider a single pixel $\boldsymbol{u}$ with the L2 loss, ${L} = ||C - C_{gt}||^2$, where ${C}$ and $C_{gt}$ denote the rendered color and the ground truth color at pixel $\boldsymbol{u}$, respectively. The gradient of the loss can be computed using the chain rule, 
	\begin{equation}\label{eq:4}
		\frac{{\partial}{L}}{{\partial}\gamma_i}=2({C}-{C_{gt}}) \frac{\partial{C}}{{\partial}\gamma_i},\gamma_i\in\Gamma_i
	\end{equation}
	By integrating along the viewing ray $l$ associated with pixel $\boldsymbol{u}$ and considering $N$ ordered Gaussians, the equation can be expanded as,
	\begin{equation}\label{eq:5}
		\frac{\partial{C}}{{\partial}\gamma_i}=\sum_{k \in N}\frac{{\partial}(T_k C_k \operatorname{Sig} (o_k){\int_{\mathbf{x} \in l}}g_k(\mathbf{x})d\mathbf{x})}{{\partial}\gamma_i}
	\end{equation}
	where $T_k = \prod_{j=1}^{k-1} (1 - \alpha_j)$ denotes the transmittance.

	As demonstrated in Appendix A.1, the magnitude of the positional gradient is significantly greater than that of the other parameter gradients when the scale of the Gaussian is small.
	\begin{equation}\label{eq:6}
		\frac{{\partial}L}{{\partial}\boldsymbol{\mu}_i}\gg\frac{{\partial}L}{{\partial}\mathbf{c}_i}\sim\frac{{\partial}L}{{\partial}o_i}\sim\frac{{\partial}L}{{\partial}\mathbf{s}_i}\sim\frac{{\partial}L}{{\partial}(\mathbf{q}_i{\cdot}r_{q,i})}
	\end{equation}
	where $\sim$ denotes asymptotic equivalence, and $r_{q,i}$ denotes the update direction of $q_i$, which is governed by the definition of the quaternion.
	

	
	
	\begin{figure*}[h]
		\centering
		\includegraphics[width=\linewidth]{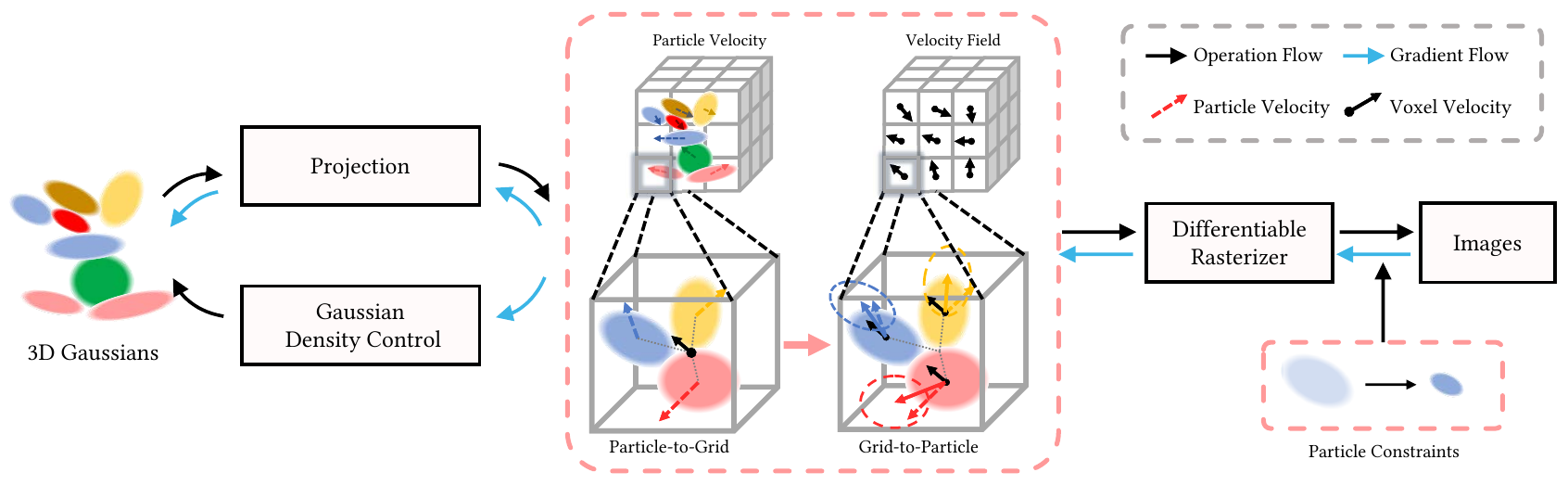}
		\caption{{\bfseries Overview of the proposed PDEO.} 3D Gaussians are initialized by COLMAP \cite{RN19}. We formulate the 3DGS optimization procedure as a Partial Differential Equation (PDE) and introduce a viscosity term to achieve stable optimization. Specifically, we employ the Material Point Method (MPM) to solve the PDE by Particle-to-Grid (P2G) and Grid-to-Particle (G2P). The velocity field is constructed to store the excess velocity of Gaussians and gradually release it to ensure the stability of Gaussian motion. In addition, we propose explicit particle constraints to enforce small-scale, high-confidence Gaussians in accordance with the particle hypothesis}
		\Description{}
		\label{fig3}
	\end{figure*}
	
	\section{Method}
	
	In this paper, we propose a new plug-and-play optimization framework, called PDEO, which leverages PDEs to enhance the rendering and reconstruction quality of 3DGS-based methods. An overview of our framework is shown in Fig. \ref{fig3}. In Section 4.1, we first establish the PDE formulation for the 3DGS optimization procedure and introduce a viscous term to enhance the stability of the optimization. Secondly, we employ the MPM to solve the PDE by Particle-to-Grid (P2G) and Grid-to-Particle (G2P) strategies in Section 4.2. Finally, we propose explicit particle constraints to enforce small-scale, high-confidence Gaussians in accordance with the particle hypothesis in Section 4.3.
	
	\subsection{PDE based 3DGS Optimization}
	
	In this section, we establish the PDE formulation for the 3DGS optimization procedure, which allows us control 3DGS optimization by explicitly modifying the PDE. 
	
	\subsubsection{Formulation}
	In a PDE, time represents the sequence of the attribute update process, allowing the system state to transition to the next state by changing attributes, similar to the iteration steps in 3DGS optimization. Thus, the attributes of Gaussians in the update process are the functions of time $t$. For the original 3DGS, the optimization procedure can be expressed as: $\boldsymbol{\mu}_{i}^{t+1}=\boldsymbol{\mu}_{i}^{t}+\sigma\frac{\partial{L^t}}{\partial{\boldsymbol{\mu}_{i}^{t}}}$, where $\sigma$ is the learning rate and $\boldsymbol{\mu}_{i}^{t}$ is the position of Gaussians $g_i$ at time $t$. We define the discrete velocity $\boldsymbol{v}_{i}^{t}$ of Gaussians $i$ at time $t$ as $\boldsymbol{v}_{i}^{t}=\boldsymbol{\mu}_{i}^{t+1}-\boldsymbol{\mu}_{i}^{t}$. Thus the velocity equation in continuous form is:
	
	\begin{equation}\label{eq:9}
		\boldsymbol{v}_{i}^{t}=\sigma\frac{\partial{L^t}}{\partial{\boldsymbol{\mu}_{i}^{t}}}
	\end{equation}
	
	Then, we calculate the partial derivatives of the equation with time $t$:
	
	\begin{equation}\label{eq:10}
		\frac{d\boldsymbol{v}_{i}^{t}}{dt}=\sigma\triangledown\frac{d{L^t}}{dt}=\sigma\frac{d}{dt}(\frac{\partial{L^t}}{\partial{\boldsymbol{\mu}_{i}^{t}}})=\sigma\sum_{\gamma_i^t{\in}\Gamma_i^t}\sigma\frac{\partial{L^t}}{\partial{{\gamma}_{i}^{t}}}\cdot\frac{\partial}{\partial{{\gamma}_{i}^{t}}}(\frac{\partial{L^t}}{\partial{\boldsymbol{\mu}_{i}^{t}}})
	\end{equation}
	
	where $\triangledown$ is the differential operator on position $\boldsymbol{\mu}_{i}^{t}$, $\gamma_i^t$ represents the attribute of $g_i$ in the attribute set $\Gamma_i^t=\{\boldsymbol{\mu}_i^t,\mathbf{c}_i^t,o_i^t,\mathbf{s}_i^t,\mathbf{q}_i^t\}$. According the definition of the time derivative and the Newton–Leibniz formula, the final motion equation is defined as:
	
	\begin{equation}\label{eq:11}
		\frac{d\boldsymbol{v}_{i}^{t}}{dt}=\frac{\partial\boldsymbol{v}_{i}^{t}}{{\partial}t}+\boldsymbol{v}_{i}^{t}\cdot\triangledown\boldsymbol{v}_{i}^{t}=\frac{\sigma^2}{2}\sum_{\gamma_i^t{\in}\Gamma_i^t}\triangledown(\frac{\partial{L^t}}{\partial{{\gamma}_{i}^{t}}})^2
	\end{equation}
	
	\subsubsection{Viscous Term}
	Unlike 3DGS optimization, particle position updating is stable and controllable during fluid simulation, which is attributed to the viscous term \cite{RN14} in the motion equations. 
	
	\begin{equation}\label{eq:12}
		\frac{{\partial}\boldsymbol{v}}{{\partial}t}+\boldsymbol{v}\cdot\triangledown\boldsymbol{v}+\frac{1}{\rho}\triangledown{p}=\boldsymbol{F}+\upsilon\triangledown\cdot\triangledown\boldsymbol{v}
	\end{equation}
	where $\triangledown\boldsymbol{v}=0$, $t$ is time, $\rho$ is density, $p$ is pressure, $\upsilon$ is viscosity, $\boldsymbol{F}$ is gravity acceleration, and $\boldsymbol{v}$ is the velocity of the fluid field, which is equal to the derivative of the particle position $\boldsymbol{\mu}$, i.e. $\boldsymbol{v}=\frac{\partial{\boldsymbol{\mu}}}{\partial{t}}$. The viscous term $\upsilon\triangledown\cdot\triangledown\boldsymbol{v}$ essentially imparts an acceleration to the particles in the system, directing them towards the average velocity of their surroundings, which can be equivalently interpreted as mixing the velocity of the particles with the average velocity of the surrounding particles.
	
	Inspired by fluid simulation \cite{RN14}, we introduce a viscous term into the 3DGS optimization procedure. Therefore, we rewrite Eq.\ref{eq:11} as:
	\begin{equation}\label{eq:14}
		\frac{d\boldsymbol{v}_{i}^{t}}{dt}=\frac{\partial\boldsymbol{v}_{i}^{t}}{{\partial}t}+\boldsymbol{v}_{i}^{t}\cdot\triangledown\boldsymbol{v}_{i}^{t}=\frac{\sigma^2}{2}\sum_{\gamma_i^t{\in}\Gamma_i^t}\triangledown(\frac{\partial{L^t}}{\partial{{\gamma}_{i}^{t}}})^2+(1-\lambda_g)\triangledown\cdot\triangledown\boldsymbol{v}_i^t
	\end{equation}
	where $\lambda_g$ is the weighting coefficient. Following the fundamental tenet of PDE, when $L$ is equal to zero, the energy of $\boldsymbol{v}$ diminishes in a gradual manner with respect to $t$ and ultimately approaches zero. Thus, introducing the viscosity does not change the solution of the equation as $t$ tends to infinity, which is the theoretical result of the 3DGS optimization.
	
	To this end, the discrete solution can be computed as:
	\begin{equation}\label{eq:15}
		\boldsymbol{\mu}_{i}^{t+1}=\boldsymbol{\mu}_{i}^{t}+\sigma\frac{\partial{L^t}}{\partial{\boldsymbol{\mu}_{i}^{t}}}+\frac{1-\lambda_g}{\left\vert N_i \right\vert}\sum_{j{\in}N_i}(\boldsymbol{v}_j^t-\boldsymbol{v}_i^t)
	\end{equation}
	where $N_i$ is the neighbour set of Gaussian $g_i$.
	
	
	\subsection{MPM based Solution}
	
	In this section, we present numerical simulations of the 3DGS optimization procedure according to the discretization form of Eq.\ref{eq:15}. We can approximate the equation as:
	\begin{equation}\label{eq:16}
		\boldsymbol{\mu}_{i}^{t+1}=\boldsymbol{\mu}_{i}^{t}+\sigma\frac{\partial{L^t}}{\partial{\boldsymbol{\mu}_{i}^{t}}}+\frac{1-\lambda_g}{\left\vert N_i \right\vert}\sum_{j{\in}N_i}(\frac{\partial{L^t}}{\partial{\boldsymbol{\mu}_{j}^{t}}}-\frac{\partial{L^t}}{\partial{\boldsymbol{\mu}_{i}^{t}}})
	\end{equation}
	Since calculating the motion of each 3D Gaussian based on its neighbour is computationally expensive after introducing the viscous term, we treat 3D Gaussians as particles and employ the MPM to solve this problem. Specifically, we incorporate the Particle-to-Grid (P2G) and Grid-to-Particle (G2P) strategies into the 3DGS optimization procedure, suppressing particle motion while providing additional motion guidance to solve the motion equation. We construct a velocity field by dividing the scene space into voxel grids. Particles can update motion by storing excess velocity in the voxel grids and gaining additional velocity from the voxel grids. Therefore, particles are effectively regulated using local information for the velocity field, thereby introducing the viscous term into the optimization procedure.

	\begin{table*}
		\belowrulesep=0pt
		\aboverulesep=0pt
		\caption{{Quantitative results on Mip-NeRF 360 \cite{RN20}, Tanks$\&$Temples \cite{RN72} and Scanet++ \cite{RN73} for Novel view synthesis.} The best results are highlighted in bold. PDEO consistently improves the performance.}
		\centering
		\label{tab1}
		\begin{tabular}{p{0.12\textwidth}|p{0.04\textwidth}p{0.04\textwidth}p{0.04\textwidth}p{0.03\textwidth}p{0.03\textwidth}|p{0.04\textwidth}p{0.04\textwidth}p{0.04\textwidth}p{0.03\textwidth}p{0.03\textwidth}|p{0.04\textwidth}p{0.04\textwidth}p{0.04\textwidth}p{0.03\textwidth}p{0.03\textwidth}}
			\toprule
			Dataset&\multicolumn{5}{c|}{Mip-NeRF360\cite{RN20}}&\multicolumn{5}{c|}{Tanks$\&$Temples\cite{RN72}}&\multicolumn{5}{c}{Scanet$++$\cite{RN73}}\\
			\toprule
			Method&$PSNR{\uparrow}$&$SSIM{\uparrow}$&$LPIPS{\downarrow}$&$Mem{\downarrow}$&$FPS{\uparrow}$&$PSNR{\uparrow}$&$SSIM{\uparrow}$&$LPIPS{\downarrow}$&$Mem{\downarrow}$&$FPS{\uparrow}$&$PSNR{\uparrow}$&$SSIM{\uparrow}$&$LPIPS{\downarrow}$&$Mem{\downarrow}$&$FPS{\uparrow}$\\
			\midrule
			3DGS&27.77&0.827&0.244&295&163.1&21.63&{0.768}&{0.322}&299&44.3&27.83&0.911&0.185&192&74.2\\
			GES&27.71&0.844&0.224&369&106.3&21.59&{0.768}&0.330&162&64.1&27.86&0.912&0.190&94.1&97.9\\
			AbaGS&27.81&0.850&0.207&804&125.1&21.37&0.755&0.326&340&40.0&27.67&0.907&0.185&121&101.8\\
			MipGS&27.98&0.858&0.213&303&108.5&20.98&0.757&0.326&357&52.1&27.80&{0.913}&0.177&224&135.2\\
			2DGS&27.42&0.841&0.228&476&42.3&21.02&0.756&0.357&188&21.8&27.91&0.911&0.196&107&36.8\\
			RaDeGS&28.03&{0.866}&{0.198}&536&118.6&20.80&0.750&0.345&239&57.5&27.97&0.911&{0.180}&165&103.4\\
			MCMC&{27.91}&0.845&0.186&714&40.4&21.03&0.744&0.318&691&55.7&28.01&0.918&0.182&470.3&52.5\\
			SpecGS&27.96&0.866&{\bfseries 0.173}&1147&7.9&21.02&0.751&0.322&498&19.7&27.89&0.912&0.195&159&56.1\\
			\midrule
			3DGS+PDEO&27.78&0.831&0.242&186&{\bfseries 225.5}&21.89&0.768&0.320&125&146.9&27.87&0.911&0.190&66.7&260.0\\
			GES+PDEO&27.99&0.834&0.232&133&166.1&22.08&0.768&0.325&97.0&{\bfseries 176.0}&27.92&0.911&0.192&53.5&{\bfseries 283.0}\\
			MipGS+PDEO&28.08&0.870&0.211&137&108.5&22.12&0.761&0.320&{\bfseries 79.0}&148.5&27.91&0.913&{\bfseries 0.169}&{\bfseries 48.5}&254.5\\
			2DGS+PDEO&27.42&0.832&0.273&{\bfseries 63.8}&94.5&21.03&0.749&0.363&100&64.6&27.93&0.911&0.195&102&81.7\\
			RaDeGS+PDEO&{28.16}&0.852&0.213&{187}&171.1&{22.61}&{0.768}&0.332&95.1&118.4&{28.06}&0.911&0.189&{65.0}&227.9\\
			MCMC+PDEO&{28.12}&0.833&0.213&198&73.3&{\bfseries 22.77}&{\bfseries 0.780}&{\bfseries 0.295}&210&73.9&28.23&{\bfseries 0.919}&0.182&212&86.9\\
			SpecGS+PDEO&{\bfseries 28.81}&{\bfseries 0.875}&{\bfseries 0.173}&99.6&65.4&22.16&{\bfseries 0.780}&{0.316}&345&28.2&{\bfseries 28.10}&{\bfseries 0.919}&0.185&115&66.8\\
			\midrule
			3DGS(rander)&{26.61}&0.764&0.318&{258}&78.8&{20.84}&{0.734}&0.380&261&66.1&{27.55}&0.908&0.202&{164.9}&92.4\\
			+PDEO(rander)&{27.75}&0.825&0.233&{89.4}&122.3&{21.71}&{0.745}&0.361&101&71.5&{27.64}&0.908&0.199&{59.5}&138.0\\
			\midrule
			MCMC(rander)&{27.62}&0.832&0.203&473&55.6&21.00&0.735&0.333&469&35.8&27.92&0.918&0.187&354&60.7\\
			+PDEO(rander)&{27.85}&0.861&0.187&189&85.9&22.64&0.771&0.321&187&54.6&27.98&0.918&0.182&98.7&63.7\\
			\bottomrule
		\end{tabular}
	\end{table*}
	
	\begin{table*}
		\belowrulesep=0pt
		\aboverulesep=0pt
		\caption{{Quantitative results on the DTU Dataset \cite{RN74} for surface reconstruction.} We report the Chamfer Distance error of different methods. The best results are highlighted in bold. PDEO consistently improves the performance.}
		\label{tab2}
		\begin{tabular}{l|ccccccccccccccc|c}
			\toprule
			Method&24&37&40&55&63&65&69&83&97&105&106&110&114&118&122&Mean\\
			\midrule
			NeRF&1.90&1.60&1.85&0.58&0.81&2.28&1.27&1.47&1.67&2.05&1.07&0.88&1.06&1.15&0.96&1.37\\
			NeuS&1.00&1.37&0.93&0.43&1.10&{\bfseries 0.65}&{\bfseries 0.57}&1.48&{\bfseries 1.09}&0.83&{\bfseries 0.52}&1.20&{\bfseries 0.35}&0.49&0.54&0.84\\
			3DGS&1.62&1.25&1.41&1.13&2.57&2.10&1.39&1.97&1.82&1.34&1.41&1.90&1.10&1.14&1.29&1.56\\
			SuGaR&1.47&1.33&1.13&0.61&2.25&1.71&1.15&1.63&1.62&1.07&0.79&2.45&0.98&0.88&0.79&1.32\\
			GOF&0.50&0.82&0.37&{\bfseries 0.37}&1.12&0.74&0.73&{\bfseries 1.18}&1.29&0.68&0.77&0.90&0.42&0.66&0.49&0.74\\
			2DGS&0.60&0.92&0.79&0.37&1.24&1.13&0.87&1.40&1.27&0.86&0.73&1.33&0.44&0.98&0.60&0.90\\
			RaDeGS&0.46&0.78&{\bfseries 0.36}&0.39&0.81&0.77&0.76&1.19&1.24&0.63&0.70&0.87&0.36&0.69&0.48&0.70\\
			\midrule
			3DGS+PDEO&1.48&1.01&1.11&0.59&2.35&1.75&1.07&1.69&1.77&0.97&1.03&1.97&1.13&1.10&1.20&1.34\\
			2DGS+PDEO&0.59&0.90&0.70&0.39&0.89&0.86&0.82&1.31&1.29&0.74&0.73&1.43
			&0.44&0.72&0.48&0.82\\
			RaDeGS+PDEO&{\bfseries 0.45}&{\bfseries 0.77}&{\bfseries 0.36}&{\bfseries 0.37}&{\bfseries 0.73}&0.75&0.75&{\bfseries 1.18}&1.16&{\bfseries 0.59}&0.67&{\bfseries 0.84}&0.38&{\bfseries 0.68}&{\bfseries 0.47}&{\bfseries 0.68}\\
			\bottomrule
		\end{tabular}
	\end{table*}
	
	\subsubsection{Particle-to-Grid}
	The P2G process constructs a grid which stores the excess velocity of particles in the voxel grids. As mentioned above, the position of the Gaussian $g_i$ is updated by ${\triangle}\boldsymbol{\mu}_i^t={\partial}L^t/{\partial}\boldsymbol{\mu}_i^t$, which is computed from the gradient of the loss. Smaller-scale Gaussians are more prone to positional mutations, which leads to instability in the optimization procedure. Thus, a reasonable reduction in velocity would be an optimization benefit. Specifically, we employ the P2G process to attenuate the particle velocity ${\triangle}\boldsymbol{\mu}_i^t$, while also preserving the motion characteristics of the particles. We store the excess velocity of the particle $g_i$ into the voxel grid $V_n$ at step $t$,
	\begin{equation}\label{eq:17}
		\boldsymbol{v}_n^{t+1}=\lambda_g\boldsymbol{v}_n^t+(1-\lambda_g)\triangle{\boldsymbol{v}_n^t}=\lambda_g\boldsymbol{v}_n^t+\frac{1-\lambda_g}{{\mid}R_n^t\mid}\sum_{g_i{\in}R_n^t}{\triangle}\boldsymbol{\mu}_i^t
	\end{equation}
	where $R_n^t$ belongs to $R^t=\{R_0^t,….,R_N^t\}$ is the set of particles contained within the voxel grid $V_n$, $\boldsymbol{v}_n^t$ is the voxel velocity saved in $V_n$, and $\lambda_g$ is weighting coefficient. We show that the selection of $\lambda_g$ has no impact on the total gradient in the Appendix A.2.
	
	\subsubsection{Grid-to-Particle}
	The grid not only suppresses particle velocity but also provides additional motion guidance for the particles. Since the velocity field represents the average motion tendency of particles in the voxel grid, the voxel velocity is then used to guide the motion of the particles:
	\begin{equation}\label{eq:18}
		{\triangle}\hat{\boldsymbol{\mu}}_i^t=\lambda_p{\triangle}\boldsymbol{\mu}_i^t+(1-\lambda_p)\boldsymbol{v}_n^t, \boldsymbol{\mu}_{i}^{t+1}=\boldsymbol{\mu}_{i}^{t}+{\triangle}\hat{\boldsymbol{\mu}}_i^t
	\end{equation}
	where the particle velocity is suppressed by the coefficient $\lambda_p$ and the ${\triangle}\hat{\boldsymbol{\mu}}_i^t$ is the updated velocity. The updated velocity represents the most likely direction of position optimization for the particles. The velocities of the different particles interact with each other, thereby cancelling out abrupt changes in position attributes across different directions while receiving additional velocity guidance from the voxel velocity. Consequently, the variation of the position gradient is successfully guided by the viscosity term.

	\subsection{Particle Constraints}
	\subsubsection{Scale Loss} In PDE, particles are scale-free attributes. Conversely, Gaussian functions with large scales can occupy a large space, which is contrary to the assumptions of PDE systems. Therefore, we introduce scale constraints for 3D Gaussians:
	\begin{equation}\label{eq:20}
		L_{s}=\frac{1}{\left\vert G_k \right\vert}{\sum_{g_i\in{G_k}}}max(s^*-\beta,0)
	\end{equation}
	where $s^*$ means the largest scale of $g_i$, $G_k$ is the set of 3D Gaussians which is visible in viewpoint $k$, and $\beta$ is the margin for the scale. This loss punishes the large scales of Gaussians. The small-scale Gaussians ensure the ability to capture high-frequency details.
	
	\subsubsection{Confidence Loss.} Since Gaussians are described as particles in the PDE, it is necessary to avoid semi-transparent Gaussians. Therefore, we propose a confidence loss to ensure the high confidence of Gaussians, satisfying the particle hypothesis, which corresponds to the opacity of the Gaussian,
	\begin{equation}\label{eq:21}
		L_{t}=\frac{1}{G_k}{\left\vert G_k \right\vert}{\mid}o_i-\lfloor1.99o_i\rfloor\mid_2^2
	\end{equation}
	where $\lfloor\cdot\rfloor$ denotes the floor operator and $o_i$ denotes the opacity. 
	
	\subsubsection{Gaussian Densification}
	
	Gaussian densification is used in 3DGS to clone and split new Gaussians to cover empty space, thus precisely representing underlying scenes. The original 3DGS averages the positional gradient of the view-space position to determine whether to perform densification. In our approach, the velocity field is also used to guide the process of cloning and splitting. Specifically, we calculate the cosine similarity measure between particle velocity $\triangle\boldsymbol{\mu}i$ and voxel velocity $\boldsymbol{v}_n$ to decide whether to perform the densify operation. Densify for Gaussian $g_i$ is performed when it satisfies $\cos(\triangle\boldsymbol{\mu}_i,\boldsymbol{v}_n)>\theta_p$, where $cos(\cdot)$ refers to cosine similarity, and $\theta_p$ denotes the cosine threshold.

	\section{Experiments}
	
	\subsection{Experimental Setup}
	
	\,\,\,\,\, \textit{Datasets.} In our experiments, we evaluated the proposed PDE-GS across a diverse range of real-world scenes to test its effectiveness in rendering and reconstruction. For novel view synthesis, we use 17 scenes from various datasets: 6 scenes from Mip-nerf360 dataset \cite{RN20}, 7 scenes from Tanks $\&$ Temples dataset \cite{RN72}, and 4 scenes from ScanNet$++$ dataset \cite{RN73}. For surface reconstruction, we conduct the experiments on 15 scenes from the DTU \cite{RN74} and 7 scenes from Tanks $\&$ Temples dataset \cite{RN72}. These scenes contain both bounded indoor and unbounded outdoor environments, enabling a comprehensive evaluation.
	
	\textit{Implementation.} To achieve high-quality rendering and reconstruction performance, our PDEO can be easily integrated into existing 3DGS-based methods, such as MipGS \cite{RN60} or 2DGS \cite{RN56}, for the tasks of novel view synthesis and surface reconstruction. To ensure consistent evaluation, we use the default parameters of the original methods. We set $\lambda_g=0.8$, $\lambda_p=0.8$, $\psi=0.2$, $\theta_p=120^\circ$, $\beta=0.6$, $\omega_t=0.04$, $\omega_s=0.04$ and $\tau$ increasing from $1$ to $2.5$ with iteration gradually. All our experiments are conducted on a single V100 GPU. 
	
	\textit{Metrics.} To evaluate the rendering quality, we report {PSNR}, {SSIM}, and {LPIPS} to measure the performance of each dataset. To evaluate the reconstruction quality, we report the {Chamfer Disrance (CD)} on DTU dataset \cite{RN74} and the {F1-score} on Tanks $\&$ Temples dataset \cite{RN72}.

	\subsection{Comparison}
	
	\subsubsection{Novel View Synthesis}
	
	We integrate the proposed PDEO into state-of-the-art 3DGS-based methods for novel view synthesis, and compare it with 3DGS \cite{RN5}, GES \cite{RN32}, AbsGS \cite{RN30}, MipGS \cite{RN60}, 2DGS \cite{RN56}, RaDeGS \cite{RN15}, 3DGS MCMC and SpecGS \cite{yang2024spec}.
	
\begin{figure*}[h]
		\centering
		\includegraphics[width=0.93\linewidth]{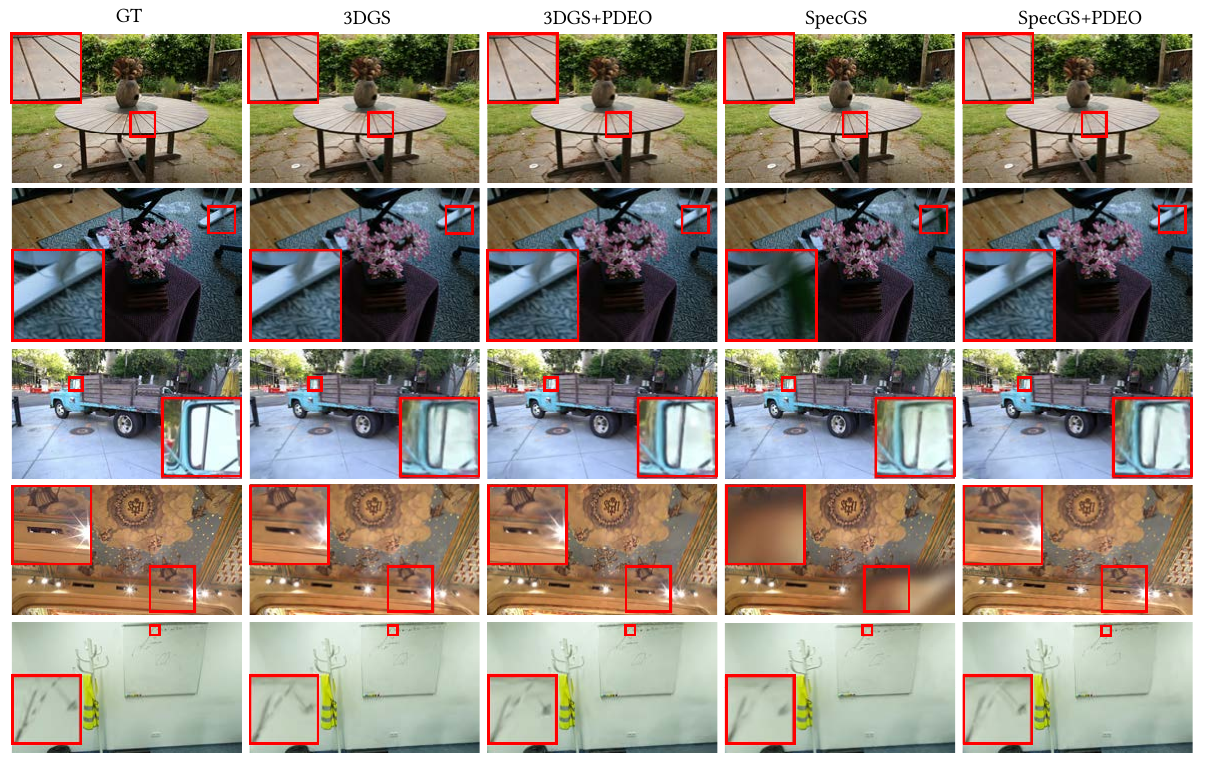}
		\caption{Qualitative comparisons of different methods on scenes from Mip-NeRF360 \cite{RN20} and Tanks$\&$Temples \cite{RN72} and Scanet$++$\cite{RN73} datasets for novel view synthesis. PEDO significantly reduces artifacts and floaters while improving rendering quality.}
		\Description{}
		\label{fig4}
	\end{figure*}

\begin{figure*}[h]
		\centering
		\includegraphics[width=0.93\linewidth]{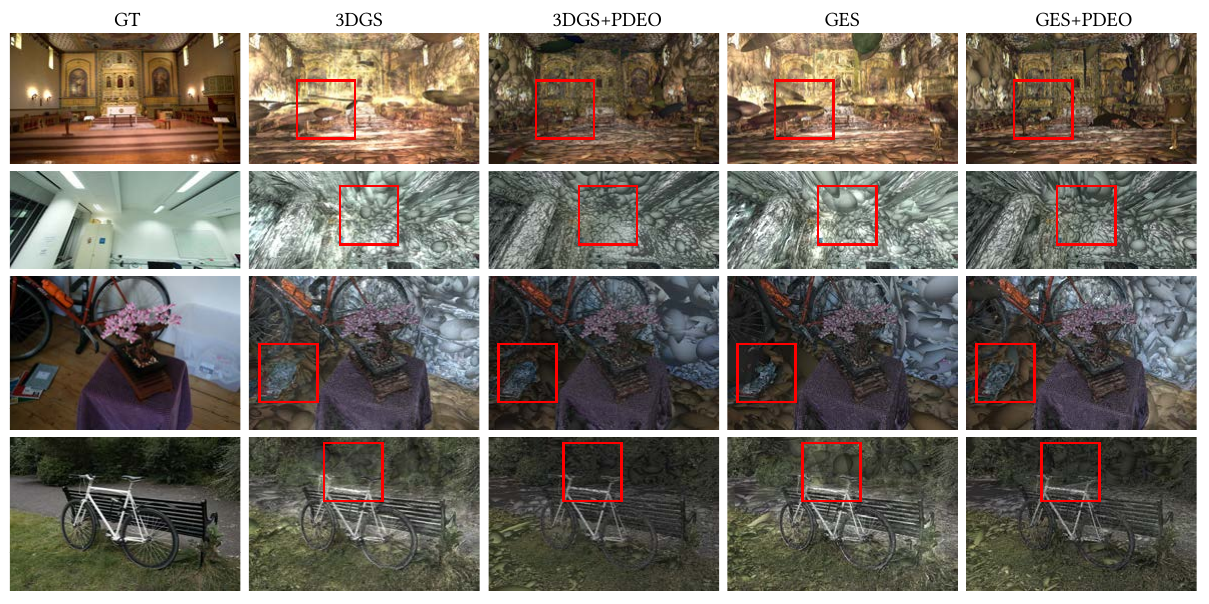}
		\caption{{ Visualization of Gaussian ellipsoids.} PEDO eliminates floater Gaussians and recovers fine geometric details.}
		\Description{}
		\label{fig5}
	\end{figure*}

	We report the quantitative results in Table \ref{tab1}. PDEO consistently improves the performance of the original methods in terms of PSNR, SSIM, and LPIPS. We can see that SpecGS+PDEO achieves the best performance. The quantitative results are shown in Fig. \ref{fig4}, demonstrating that PDEO significantly reduces artifacts and floaters while improving rendering quality. For a clear comparison, we also provide visualizations of Gaussian ellipsoids in Fig. \ref{fig5}. Overall, the proposed PDEO significantly enhances 3DGS-based methods while also improving memory efficiency.

	\begin{table}
		\belowrulesep=0pt
		\aboverulesep=0pt
		\caption{{ Quantitative results on Tanks$\&$temples \cite{RN72} for surface reconstruction.} We report the F1-score of different methods. The best results are highlighted in bold. RaDeGS+PDEO achieves the best F1-score among all compared methods.}
		\label{tab:freq}
		\begin{tabular}{l|cccc}
			\toprule
			Method&2DGS&RaDeGS&SuGaR&RaDeGS+PDEO\\
			\midrule
			Barn&0.387&0.470&0.171&{\bfseries 0.588}\\
			Caterpillar&0.210&0.255&0.129&{\bfseries 0.343}\\
			Courthouse&0.126&0.100&0.084&{\bfseries 0.128}\\
			Ignatius&0.517&0.668&0.351&{\bfseries 0.780}\\
			Meetingroom&0.250&0.240&0.180&{\bfseries 0.610}\\
			Truck&0.379&0.462&0.225&{\bfseries 0.591}\\
			Church&0.054&0.018&0.035&{\bfseries 0.078}\\
			\midrule
			Mean&0.275&0.316&0.168&{\bfseries 0.445}\\
			\bottomrule
		\end{tabular}
		\label{tab3}
	\end{table}

	\subsubsection{Surface Reconstruction}
\begin{figure*}[h]
		\centering
		\includegraphics[width=\linewidth]{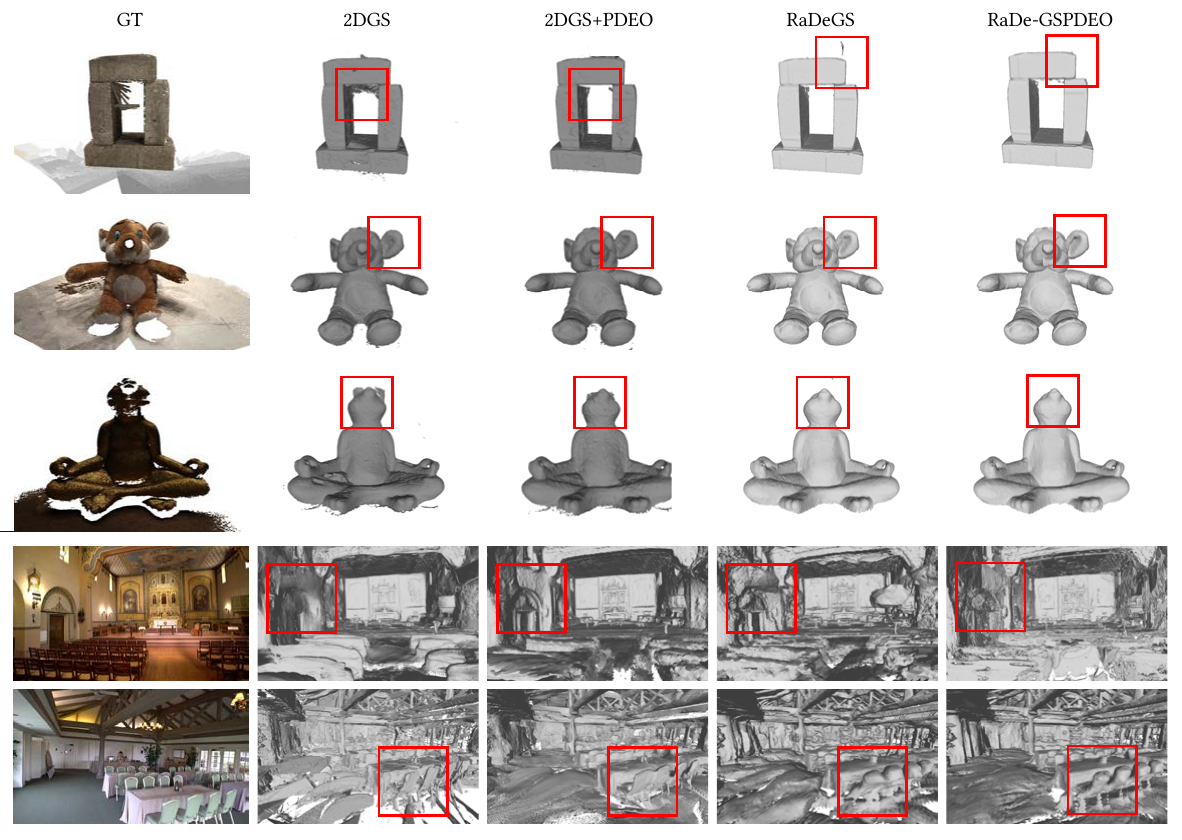}
		\caption{{ Qualitative comparisons of different methods on scenes from Tanks$\&$Temples \cite{RN72} datasets for surface reconstruction.} PEDO improves the quality of the reconstruction.}
		\Description{}
		\label{fig6}
	\end{figure*}

	PDEO is integrated with state-of-the-art 3DGS-based methods for surface reconstruction and compared with 2DGS \cite{RN56}, RaDeGS \cite{RN15}, and SuGaR \cite{RN59}. As shown in Table \ref{tab2} and Table \ref{tab3}, PDEO consistently enhances 3DGS-based methods on the DTU dataset in terms of CD error, and on the Tanks \& Temples dataset in terms of F1-score. RaDeGS+PDEO achieves qualitatively better reconstructions with more accurate and smoother geometry, as shown in Fig. \ref{fig6}. This demonstrates that PDEO can remove floaters and preserve geometric details to improving reconstruction quality.

	\begin{table}
		\belowrulesep=0pt
		\aboverulesep=0pt
		\caption{{Ablation study on Mip-NeRF360 Dataset \cite{RN20} for novel view synthesis.} We study the influence of each component in our method on the rendering quality and memory usage in Mip-NeRF360 Dataset \cite{RN20}.}
		\label{tab4}
		\begin{tabular}{l|cccc}
			\toprule
			\multicolumn{1}{l|}{Method}&$PSNR{\uparrow}$&$SSIM{\uparrow}$&$LPIPS{\downarrow}$&$Mem{\downarrow}$\\
			\toprule
			\multicolumn{1}{l|}{Baseline}&27.71&0.844&0.224&369\\
			\multicolumn{1}{l|}{w/o P2G and G2P}&27.51&0.830&0.227&240\\
			\multicolumn{1}{l|}{w/o Our Densification}&27.96&0.834&0.230&136\\
			\multicolumn{1}{l|}{w/o Scale Loss}&27.75&0.831&0.236&{\bfseries 132}\\
			\multicolumn{1}{l|}{w/o Confidence Loss}&27.87&{\bfseries0.845}&{\bfseries 0.219}&177\\
			\multicolumn{1}{l|}{Full}&{\bfseries 27.99}&0.834&0.232&133\\
			\multicolumn{1}{l|}{Full($\lambda_g$=0.5)}&{27.84}&0.835&0.235&151\\
			\multicolumn{1}{l|}{Full($\lambda_g$=0.9)}&{27.66}&0.834&0.233&160\\
			\multicolumn{1}{l|}{Full($\lambda_p$=0.5)}&{27.69}&0.835&0.237&143\\
			\multicolumn{1}{l|}{Full($\lambda_p$=0.9)}&{27.56}&0.833&0.229&207\\
			\bottomrule
		\end{tabular}
	\end{table}
	
	\subsection{Ablation Studies}
	
	In this section, we conduct ablation experiments to study the effectiveness of each component of PDEO. We conduct experiments on Mip-NeRF360 dataset \cite{RN20} for novel view synthesis. The quantitative results of the ablations are reported in Table 4 and GES \cite{RN32} is used as the baseline.
	
\begin{figure*}[t]
		\centering
		\includegraphics[width=\linewidth]{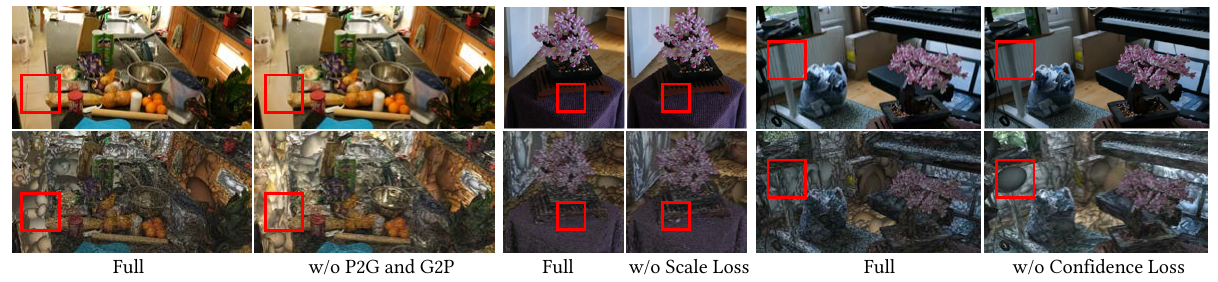}
		\caption{{ Ablation of P2G and G2P, the Scale Loss and Confidence Loss}}
		\Description{}
		\label{fig7}
	\end{figure*}

	\textit{Effects of P2G and G2P.}
	In Table \ref{tab4}, we examine the impact of P2G and G2P. The absence of this strategy leads to a significant decline in rendering quality, which leads to a decrease in rendering quality. Our approach introduces the viscosity term into the optimization procedure using P2G and G2P strategies, which can ensure stable optimization of Gaussians while reducing memory usage. Additionally, the qualitative rendering results are illustrated in Fig. \ref{fig7}, which demonstrate that P2G and G2P help mitigate artifacts and floaters.
	
	\textit{Effects of Gaussian Densification.}
	As shown in Table \ref{tab4}, removing the Gaussian densification strategy results in a degradation of rendering quality, demonstrates that the strategy can achieve more accurate Gaussian densification to fit the details of scenes.
	
	\textit{Effects of Scale Loss and Confidence Loss.}
	We analyze the effects of scale loss and confidence loss. Table \ref{tab4} shows that with a similar amount of memory usage, there is a significant degradation in rendering quality when removing scale loss or confidence loss. Fig. \ref{fig7} evidences that scale loss helps limit the scale attribute of Gaussians, which facilitates a better reconstruction of scene details. 
	
	\section{CONCLUSION}
	
	The reconstruction of detailed features in a scene requires optimizing numerous small-scale 3D Gaussians. However, to these 3D Gaussians, the sensitivity magnitude of the positional gradient is significantly higher than that of the other parameter gradients. The unequal optimization treatment to different Gaussian attributes according to the computation of gradient magnitude leads to the unstable optimization of 3DGS. Therefore, we propose PDEO, which builds the correspondence between the 3DGS optimization and the PDE simulation, to control and guide the 3DGS optimization. Our experimental results demonstrate its effectiveness in enhancing render and reconstruction quality.
	
	\textit{Limitation.} Our method exhibits a couple of limitations. Firstly, Our method does not involve particle rotation. Future research could incorporate the influence of the spatial voxel grids on particle rotation in the MPM simulation. Secondly, although we introduce voxel grids to guide the optimization of particle direction, our approach still struggles to reliably relocate 3D Gaussians from other regions into areas with substantial gaps in point cloud initialization (e.g., regions of the scene that lack an initial point cloud). Addressing these limitation represents a promising direction for future research.

	\bibliographystyle{ACM-Reference-Format}
	\bibliography{sample-base}

\newpage
\newpage
\newpage
\clearpage
\appendix

	\section{Appendix}

	\subsection{Gaussian Gradient Sensitivity Analysis}
	
	{\bfseries Attributes in 3DGS. }For some attributes with restricted value ranges, 3DGS applies an activation function to map an unbounded attributes to a bounded value range. Below is a comparison of some attributes and their corresponding rendering properties used in 3DGS:
	\begin{itemize}
		\item{possion: }$\boldsymbol{\mu_i}\in R^3$  
		\item{color: }$\hat{\mathbf{c}}_{i,\phi}=f(\phi,\mathbf{c}_i),$
		\item{opacity: }$\hat{o}_i=\operatorname{Sig}(o_i),$  
		\item{scale: }$\hat{\mathbf{s}}_i = e^{(\mathbf{s}_{i})},$
		\item{rotation: }$\mathbf{q}_{i}\in R^4$
	\end{itemize}
	where $\boldsymbol{\mu}_i$ denotes the center position, $\mathbf{c}_i$ represents the spherical harmonic coefficients, $o_i$ is the opacity attribute, $\mathbf{s}_i$ refers to the scale attributes, and $\mathbf{q}_i$ is the quaternion representing the rotation attributes. Here, $\hat{o}_i = \operatorname{Sig}(o_i)$ denotes the opacity, where $\operatorname{Sig}(\cdot)$ represents the sigmoid function, and $\hat{\mathbf{s}}_i = e^{(\mathbf{s}_{i})} \in R^3$ denotes the scaling vector.
	
	\begin{figure}[h]
		\centering
		\includegraphics[width=\linewidth]{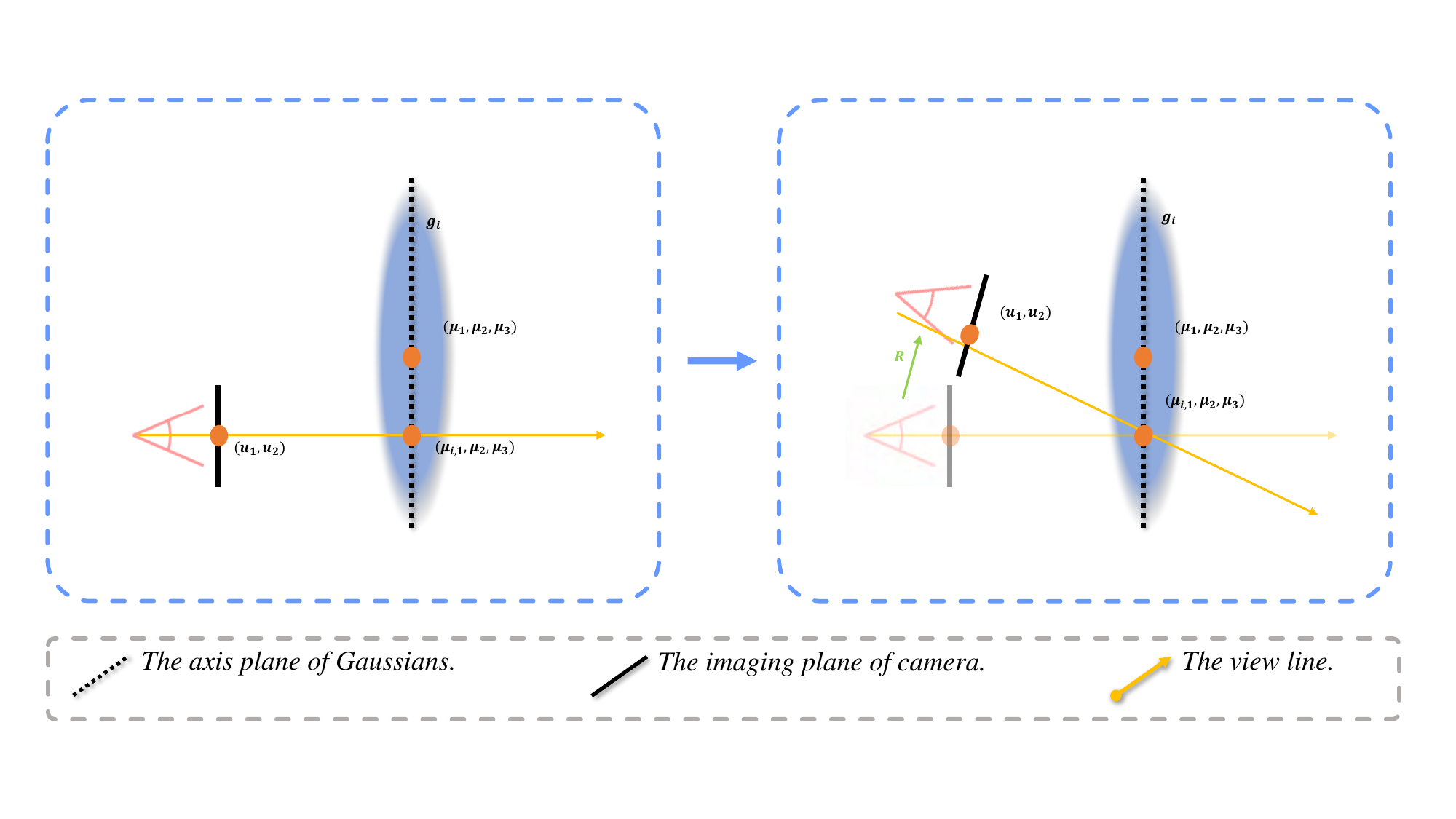}
		\caption{{Projection process of 3D Gaussians. Left.} The view line of camera is orthogonal to the axis plane of 3D Gaussian.{\bfseries Right.} The situation is the same for rotation Gaussians and rotation cameras, so we choose to rotate cameras. After a rotation R, the view line is not orthogonal to the plane.}
		\Description{}
		\label{fig}
	\end{figure}
	
	{\bfseries 3DGS to 2D splatting. }For a simple non-rotated 3D Gaussian basis function:
	$$g_i(\boldsymbol{\mu})=e^{{-\frac{1}{2}(\boldsymbol{\mu}-\boldsymbol{\mu}_i)^T\boldsymbol{\Sigma}^{-1}(\boldsymbol{\mu}-\boldsymbol{\mu}_i)}},$$
	here $\boldsymbol{\mu}=(\mu_{1},\mu_{2},\mu_{3})$ is the sampling possition and $\boldsymbol{\mu}_i=(\mu_{i,1},\mu_{i,2},\mu_{i,3})$ is the possition of the Gaussian $g_i$. If we integrate along one of the coordinate axes  $(1,0,0)$ throgh the point $(\mu_{i,1},\mu_{2},\mu_{3})$ and the corresponding pixel is $\boldsymbol{u}=(u_1,u_2)$, To simplify, we let $\boldsymbol{x_i}=\boldsymbol{\mu-\mu_i},(\boldsymbol{x_i}=(x_{i,1},x_{i,2},x_{i,3}))$. We obtain the following integral result:
	$$splat_i(\boldsymbol{u})=splat_i(u_1,u_2)={\int}g_i(\boldsymbol{\mu})d\mu_1={\int}e^{-(\frac{x_{i,1}^2}{2{\cdot}\hat{\mathbf{s}}_{i,1}^2}+\frac{x_{i,2}^2}{2{\cdot}\hat{\mathbf{s}}_{i,2}^2}+\frac{x_{i,3}^2}{2{\cdot}\hat{\mathbf{s}}_{i,3}^2})}d\mu_1$$
	{\raggedleft
		$
		$$=e^{-(\frac{x_{i,2}^2}{2{\cdot}\hat{\mathbf{s}}_{i,2}^2}+\frac{x_{i,3}^2}{2{\cdot}\hat{\mathbf{s}}_{i,3}^2})}{\cdot}{\sqrt{2\pi}}{\cdot}\hat{\mathbf{s}}_{i,1},$$
		$\par}
	where $\boldsymbol{\Sigma}^{-1}$ is a $3\cdot3$ metrix as:
	$$\begin{bmatrix}
		\frac{1}{2\hat{\mathbf{s}}_{i,1}^2} & 0 & 0 \\
		0 & \frac{1}{2\hat{\mathbf{s}}_{i,2}^2} & 0 \\
		0 & 0 & \frac{1}{2\hat{\mathbf{s}}_{i,3}^2}
	\end{bmatrix}$$
	
	This is the 2D splatting function at the pixel projected from point $(\mu_{i,1},\mu_{2},\mu_{3})$ in the absence of rotation.
	
	And when we introduce a rotation matrix $\boldsymbol{R}$, the integral of the rotated function along a line passing through point $(\mu_{i,1},\mu_{2},\mu_{3})$ and parallel to the viewing direction is equivalent to the integral of the non-rotated function along a line passing through point $(\mu_{i,1},\mu_{2},\mu_{3})$ that has been rotated by $\boldsymbol{R}$.
	
	Then We assume that the direction vector of the integration axis after rotation is $r=(r_1,r_2,r_3)$, where $r_1^2+r_2^2+r_3^2=1$. So the integral result of the rotated function 
	$$g_i(\boldsymbol{\mu})=e^{-\boldsymbol{x_i}^TR^T{\Sigma}R\boldsymbol{x_i}},$$ 
	along $(1,0,0)$, we integrate it
	$$splat_i(\boldsymbol{u})={\int}g_i(\boldsymbol{\mu})d\mu_1={\int}e^{-(\frac{(r_1{\cdot}t)^2}{2{\cdot}\hat{\mathbf{s}}_1^2}+\frac{(r_2{\cdot}t+x_{i,2})^2}{2{\cdot}\hat{\mathbf{s}}_2^2}+\frac{(r_3{\cdot}t+x_{i,3})^2}{2{\cdot}\hat{\mathbf{s}}_3^2})}dt,$$
	we simplify it to 
	$$e^{-(\frac{x_{i,2}^2}{2{\cdot}\hat{\mathbf{s}}_2^2}+\frac{x_{i,3}^2}{2{\cdot}\hat{\mathbf{s}}_3^2})}{\cdot}{\int}e^{-(\frac{r_1^2{\cdot}t^2}{2{\cdot}\hat{\mathbf{s}}_1^2}+\frac{r_2^2{\cdot}t^2}{2{\cdot}\hat{\mathbf{s}}_2^2}+\frac{r_3^2{\cdot}t^2}{2{\cdot}\hat{\mathbf{s}}_3^2}+\frac{r_2{\cdot}t{\cdot}x_{i,2}}{\hat{\mathbf{s}}_2^2}+\frac{r_3{\cdot}t{\cdot}x_{i,3}}{\hat{\mathbf{s}}_3^2})}dt,$$
	so we introduce two coefficients $A=\frac{r_1^2}{2{\cdot}\hat{\mathbf{s}}_1^2}+\frac{r_2^2}{2{\cdot}\hat{\mathbf{s}}_2^2}+\frac{r_3^2}{2{\cdot}\hat{\mathbf{s}}_3^2}$, and $B(x_{i,2},x_{i,3})=\frac{r_2{\cdot}x_{i,2}}{\hat{\mathbf{s}}_2^2}+\frac{r_3{\cdot}x_{i,3}}{\hat{\mathbf{s}}_3^2}$:
	$$e^{-(\frac{x_{i,2}^2}{2{\cdot}\hat{\mathbf{s}}_2^2}+\frac{x_{i,3}^2}{2{\cdot}\hat{\mathbf{s}}_3^2})+\frac{B(x_{i,2},x_{i,3})^2}{4{\cdot}A}}{\cdot}{\int}e^{-A(t+\frac{B(x_{i,2},x_{i,3})}{2A})^2}dt,$$
	so we can get 
	$$splat(\boldsymbol{x_i})=splat_i(\boldsymbol{u})=e^{-(\frac{x_{i,2}^2}{2{\cdot}\hat{\mathbf{s}}_2^2}+\frac{x_{i,3}^2}{2{\cdot}\hat{\mathbf{s}}_3^2})+\frac{B(x_{i,2},x_{i,3})^2}{4{\cdot}A}}{\cdot}\sqrt{\frac{{\pi}}{A}},$$
	as the splatting result of $g_i$ at the pixel $\boldsymbol{u}$.
	
	{\bfseries Renderring Gradient. }For the energy term of rendering supervision, we can write it as:
	$$L=\sum_{\boldsymbol{u}}(render(\boldsymbol{u})-gt(\boldsymbol{u}))^2,$$
	here $gt(\cdot)$ is the ground truth of the view and the $render(\boldsymbol{u})$ is the render function of Gaussian splatting which can be writen as:
	$$render(\boldsymbol{u})=\sum_{i}T_i\hat{\mathbf{c}}_i(\operatorname{Sig}(o_i){\cdot}splat(\boldsymbol{x_i})).$$
	where $\hat{\mathbf{c}}_i$ is color and $T_i=\Pi_{k=1}^{i-1}(1-\alpha_k)$ is transmittance of $g_i$, here $\alpha_k=\operatorname{Sig}(o_k){\cdot}splat(\boldsymbol{x_k})$ is opacity.
	We find
	
	$$\frac{{\partial}L}{{\partial}\gamma_i}=\sum_{\boldsymbol{u}}2(render(\boldsymbol{u})-gt(\boldsymbol{u}))\cdot\sum_{k}\frac{{\partial}(T_k\hat{\mathbf{c}}_k(\operatorname{Sig}(o_k)*splat(\boldsymbol{x_k})))}{{\partial}\gamma_i}$$
	here $k$ is also the index of gaussians, and $\gamma_i\in\{\boldsymbol{\mu}_i^t,c_i^t,o_i^t,s_i^t,q_i^t\}$ is the attributes of $g_i$. So we can only  discuss 
	$$\frac{{\partial}(T_k{\cdot}\hat{\mathbf{c}}_k{\cdot}\alpha_k)}{{\partial}\gamma_i}=\frac{{\partial}(T_k\hat{\mathbf{c}}_k(\operatorname{Sig}(o_k)*splat(\boldsymbol{x_k})))}{{\partial}\gamma_i},$$
	if we want compare the gradients of diffierent attributes. 
	
	So when $k=i$, we have:
	$$\frac{{\partial}(T_i{\cdot}\hat{\mathbf{c}}_i{\cdot}\alpha_i)}{{\partial}{\hat{\mathbf{c}}_i}}=T_i\alpha_i\frac{{\partial}\hat{\mathbf{c}}_i}{{\partial}{c_i}},$$
	$$\frac{{\partial}(T_i{\cdot}\hat{\mathbf{c}}_i{\cdot}\alpha_i)}{{\partial}o_i}=T_i\hat{\mathbf{c}}_i(1-\operatorname{Sig}(o_i))\operatorname{Sig}(o_i)splat(\boldsymbol{x_i}),$$
	$$\frac{{\partial}(T_i{\cdot}\hat{\mathbf{c}}_i{\cdot}\alpha_i)}{{\partial}\mu_{i,j}}=T_i\hat{\mathbf{c}}_i\operatorname{Sig}(o_i)(splat(\boldsymbol{x_i}))_{\mu_{i,j}},$$
	$$\frac{{\partial}(T_i{\cdot}\hat{\mathbf{c}}_i{\cdot}\alpha_i)}{{\partial}s_{i,j}}=T_i\hat{\mathbf{c}}_i\operatorname{Sig}(o_i)(splat(\boldsymbol{x_i}))_{s_{i,j}},$$
	here $(\cdot)_\gamma$ denotes the partial derivative. And when $k$ is different from $i$:
	
	$$\frac{{\partial}(T_k{\cdot}\hat{\mathbf{c}}_k{\cdot}\alpha_k)}{{\partial}{c_i}}=0,$$
	$$\frac{{\partial}(T_k{\cdot}\hat{\mathbf{c}}_k{\cdot}\alpha_k)}{{\partial}o_i}=-\frac{T_k\hat{\mathbf{c}}_k\alpha_k(1-\operatorname{Sig}(o_i))\operatorname{Sig}(o_i)splat(\boldsymbol{x_i})}{1-\alpha_i},$$
	$$\frac{{\partial}(T_k{\cdot}\hat{\mathbf{c}}_k{\cdot}\alpha_k)}{{\partial}\mu_{i,j}}=-\frac{T_k\hat{\mathbf{c}}_k\alpha_k\operatorname{Sig}(o_i)(splat(\boldsymbol{x_i}))_{\mu_{i,j}}}{1-\alpha_i},$$
	$$\frac{{\partial}(T_k{\cdot}\hat{\mathbf{c}}_k{\cdot}\alpha_k)}{{\partial}s_{i,j}}=-\frac{T_k\hat{\mathbf{c}}_k\alpha_k\operatorname{Sig}(o_i)(splat(\boldsymbol{x_i}))_{s_{i,j}}}{1-\alpha_i},$$
	where
	$$splat(\boldsymbol{x_i})=e^{-(\frac{(\mu_{i,2}-\mu_2)^2}{2{\cdot}\hat{\mathbf{s}}_{i,2}^2}+\frac{(\mu_{i,3}-\mu_3)^2}{2{\cdot}\hat{\mathbf{s}}_{i,3}^2})+\frac{B(\mu_{i,2}-\mu_2,\mu_{i,3}-\mu_3)^2}{4{\cdot}A}}{\cdot}\sqrt{\frac{{\pi}}{A}}.$$

	So if we ignore $(splat(\boldsymbol{x_i}))_{\mu_{i,j}}$ and $(splat(\boldsymbol{x_i}))_{s_{i,j}}$, we can find the remaining parts of the items in the same group are of the same magnitude. For $\alpha_i{\sim}\operatorname{Sig}(o_i)$ and $\hat{\mathbf{c}}_i{\sim}\frac{\partial{\hat{\mathbf{c}}_i}}{{\hat{\mathbf{c}}_i}}{\sim}\alpha_i{\sim}(1-\operatorname{Sig}(o_i)){\sim}splate(\boldsymbol{x_i}){\sim}1$. So we can only judge $(splat(\boldsymbol{x_i}))_{\mu_{i,j}}$ and $(splat(\boldsymbol{x_i}))_{s_{i,j}}$ to compare the gradients.
	
	Let $j=2$, then we can get
	$$(splat_i)_{\mu_{i,2}}=\frac{1}{\hat{\mathbf{s}}_{i,2}^2}splat_i{\cdot}(\frac{\frac{r_2r_3(\mu_{i,3}-\mu_3)}{\hat{\mathbf{s}}_{i,3}^2}-(\frac{r_1^2}{\hat{\mathbf{s}}_{i,1}^2}+\frac{r_3^2}{\hat{\mathbf{s}}_{i,3}^2})(\mu_{i,2}-\mu_2)}{\frac{r_1^2}{\hat{\mathbf{s}}_{i,1}^2}+\frac{r_2^2}{\hat{\mathbf{s}}_{i,2}^2}+\frac{r_3^2}{\hat{\mathbf{s}}_{i,3}^2}}),$$
	here $splat_i=splate(\boldsymbol{x_i})$.  Obviously, we have $splat_i{\sim}1$, 
	$(\mu_{i,3}-\mu_{3}){\sim}\hat{\mathbf{s}}_{i,3}$ and $ax^2+by^2\geq2\sqrt{ab}xy$, so we have $$\frac{\frac{r_2r_3(\mu_{i,3}-\mu_{3})}{\hat{\mathbf{s}}_{i,3}^2}}{\frac{r_1^2}{\hat{\mathbf{s}}_{i,1}^2}+\frac{r_2^2}{\hat{\mathbf{s}}_{i,2}^2}+\frac{r_3^2}{\hat{\mathbf{s}}_{i,3}^2}}\sim\frac{r_2r_3}{\frac{\hat{\mathbf{s}}_{i,3}r_1^2}{\hat{\mathbf{s}}_{i,1}^2}+\frac{\hat{\mathbf{s}}_{i,3}r_2^2}{\hat{\mathbf{s}}_{i,2}^2}+\frac{r_3^2}{\hat{\mathbf{s}}_{i,3}}}\leq\frac{r_2r_3}{\frac{2r_2r_3}{\hat{\mathbf{s}}_{i,2}}}{\sim}\hat{\mathbf{s}}_{i,2},$$
	and we have $(\mu_{i,2}-\mu_{2}){\sim}\hat{\mathbf{s}}_{i,2}$, so $$\frac{(\frac{r_1^2}{\hat{\mathbf{s}}_{i,1}^2}+\frac{r_3^2}{\hat{\mathbf{s}}_{i,3}^2})(\mu_{i,2}-\mu_{2})}{\frac{r_1^2}{\hat{\mathbf{s}}_{i,1}^2}+\frac{r_2^2}{\hat{\mathbf{s}}_{i,2}^2}+\frac{r_3^2}{\hat{\mathbf{s}}_{i,3}^2}}\leq(\mu_{i,2}-\mu_{2}){\sim}\hat{\mathbf{s}}_{i,2}.$$
	According to the definition of equivalence we can get
	$(splat_i)_{\mu_{i,2}}\lesssim\frac{1}{\hat{\mathbf{s}}_{i,2}},$
	and when $r_2=0$, $(splat_i)_{\mu_{i,2}}{\sim}\frac{1}{\hat{\mathbf{s}}_{i,2}}.$
	So $(splat_i)_{\mu{i,2}}{\sim}\frac{1}{\hat{\mathbf{s}}_{i,2}}$. And similarly at $j=3$, we have $(splat_i)_{\mu{i,3}}{\sim}\frac{1}{\hat{\mathbf{s}}_{i,3}}$.
	
	Similarly, we also handle $(splat_i)_{s_{i,j}}$, as before, we only need to take $j=2$, since the other value of $j$ is the same as $j=2$. Noting that $\hat{\mathbf{s}}_i=e^{s_i}$ and $(\hat{\mathbf{s}}_i)_{s_i}=\hat{\mathbf{s}}_i$.
	$$(splat_i)_{s_{i,2}}=-\frac{(\frac{r_1^2}{\hat{\mathbf{s}}_{i,1}^2}+\frac{r_3^2}{\hat{\mathbf{s}}_{i,3}^2})r_2^2(\mu_{i,2}-\mu_{2})^2}{2A^2\hat{\mathbf{s}}_{i,2}^4}+\frac{(\frac{r_1^2}{\hat{\mathbf{s}}_{i,1}^2}+\frac{r_3^2}{\hat{\mathbf{s}}_{i,3}^2})(\mu_{i,2}-\mu_{2})^2}{A\hat{\mathbf{s}}_{i,2}^2}$$
	$$-\frac{2r_2r_3(\mu_{i,2}-\mu_{2})(\mu_{i,3}-\mu_{3})(\frac{\hat{\mathbf{s}}_{i,3}^2r_1^2}{\hat{\mathbf{s}}_{i,1}^2}+r_3^2)}{\hat{\mathbf{s}}_{i,2}^2\hat{\mathbf{s}}_{i,3}^4A^2}.$$
	
	We analyze each item step by step. For $(\mu_{i,2}-\mu_{2}){\sim}\hat{\mathbf{s}}_{i,2}$,
	$$\frac{(\frac{r_1^2}{\hat{\mathbf{s}}_{i,1}^2}+\frac{r_3^2}{\hat{\mathbf{s}}_{i,3}^2})r_2^2(\mu_{i,2}-\mu_{2})^2}{2A^2\hat{\mathbf{s}}_{i,2}^4}\lesssim\frac{r_2^2}{2A\hat{\mathbf{s}}_{i,2}^2}{\sim}1,$$
	Similarly, we obtain:
	$$\frac{(\frac{r_1^2}{\hat{\mathbf{s}}_{i,1}^2}+\frac{r_3^2}{\hat{\mathbf{s}}_{i,3}^2})(\mu_{i,2}-\mu_{2})^2}{A\hat{\mathbf{s}}_{i,2}^2}{\sim}1$$
	
	The third item is slightly more complex, so we will handle it in two parts. Firstly, We will address the first part:
	$$E_1:=\frac{2r_1^2r_2r_3(\mu_{i,2}-\mu_{2})(\mu_{i,3}-\mu_{3})}{\hat{\mathbf{s}}_{i,1}^2\hat{\mathbf{s}}_{i,2}^2\hat{\mathbf{s}}_{i,3}^2A^2}.$$
	
	Note $E_1=\frac{2\hat{\mathbf{s}}_{i,2}^2\hat{\mathbf{s}}_{i,3}^2r_1^2\hat{\mathbf{s}}_{i,1}^2r_2r_3(\mu_{i,2}-\mu_{2})(\mu_{i,3}-\mu_{3})}{(\hat{\mathbf{s}}_{i,2}^2\hat{\mathbf{s}}_{i,3}^2r_1^2+\hat{\mathbf{s}}_{i,1}^2\hat{\mathbf{s}}_{i,3}^2r_2^2+\hat{\mathbf{s}}_{i,1}^2\hat{\mathbf{s}}_{i,2}^2r_3^2)^2},$
	so a natural thinking is dividing it into two parts:
	$$E_1=\frac{2\hat{\mathbf{s}}_{i,2}^2\hat{\mathbf{s}}_{i,3}^2r_1^2}{\hat{\mathbf{s}}_{i,2}^2\hat{\mathbf{s}}_{i,3}^2r_1^2+\hat{\mathbf{s}}_{i,1}^2\hat{\mathbf{s}}_{i,3}^2r_2^2+\hat{\mathbf{s}}_{i,1}^2\hat{\mathbf{s}}_{i,2}^2r_3^2}$$
	$${\cdot}\frac{\hat{\mathbf{s}}_{i,1}^2r_2r_3(\mu_{i,2}-\mu_{2})(\mu_{i,3}-\mu_{3})}{\hat{\mathbf{s}}_{i,2}^2\hat{\mathbf{s}}_{i,3}^2r_1^2+\hat{\mathbf{s}}_{i,1}^2\hat{\mathbf{s}}_{i,3}^2r_2^2+\hat{\mathbf{s}}_{i,1}^2\hat{\mathbf{s}}_{i,2}^2r_3^2}.$$
	
	For $(\mu_{i,2}-\mu_{2}){\sim}\hat{\mathbf{s}}_{i,2}$ and $(\mu_{i,3}-\mu_{3}){\sim}\hat{\mathbf{s}}_{i,3}$, we have:
	$$\frac{2\hat{\mathbf{s}}_{i,2}^2\hat{\mathbf{s}}_{i,3}^2r_1^2}{\hat{\mathbf{s}}_{i,2}^2\hat{\mathbf{s}}_{i,3}^2r_1^2+\hat{\mathbf{s}}_{i,1}^2\hat{\mathbf{s}}_{i,3}^2r_2^2+\hat{\mathbf{s}}_{i,1}^2\hat{\mathbf{s}}_{i,2}^2r_3^2}\lesssim1,$$
	$$\frac{\hat{\mathbf{s}}_{i,1}^2r_2r_3(\mu_{i,2}-\mu_{2})(\mu_{i,3}-\mu_{3})}{\hat{\mathbf{s}}_{i,2}^2\hat{\mathbf{s}}_{i,3}^2r_1^2+\hat{\mathbf{s}}_{i,1}^2\hat{\mathbf{s}}_{i,3}^2r_2^2+\hat{\mathbf{s}}_{i,1}^2\hat{\mathbf{s}}_{i,2}^2r_3^2}\lesssim1.$$
	
	Similarly, we address the second part:
	$$E_2:=\frac{2r_2r_3^3(\mu_{i,2}-\mu_{2})(\mu_{i,3}-\mu_{3})}{\hat{\mathbf{s}}_{i,2}^2\hat{\mathbf{s}}_{i,3}^4A^2}.$$
	
	Note
	$E_2=\frac{2r_2r_3(\mu_{i,3}-\mu_{3})r_3^2(\mu_{i,2}-\mu_{2})}{(\frac{\hat{\mathbf{s}}_{i,3}^2\hat{\mathbf{s}}_{i,2}r_1^2}{\hat{\mathbf{s}}_{i,1}^2}+\frac{\hat{\mathbf{s}}_{i,3}^2r_2^2}{\hat{\mathbf{s}}_{i,2}}+\hat{\mathbf{s}}_{i,2}r_3^2)^2},$ so we divide it into two parts:
	
	$$E_2=\frac{2r_2r_3(\mu_{i,3}-\mu_{3})}{\frac{\hat{\mathbf{s}}_{i,3}^2\hat{\mathbf{s}}_{i,2}r_1^2}{\hat{\mathbf{s}}_{i,1}^2}+\frac{\hat{\mathbf{s}}_{i,3}^2r_2^2}{\hat{\mathbf{s}}_{i,2}}+\hat{\mathbf{s}}_{i,2}r_3^2}{\cdot}\frac{r_3^2(\mu_{i,2}-\mu_{2})}{\frac{\hat{\mathbf{s}}_{i,3}^2\hat{\mathbf{s}}_{i,2}r_1^2}{\hat{\mathbf{s}}_{i,1}^2}+\frac{\hat{\mathbf{s}}_{i,3}^2r_2^2}{\hat{\mathbf{s}}_{i,2}}+\hat{\mathbf{s}}_{i,2}r_3^2}.$$
	
	Then we have:
	$$\frac{2r_2r_3(\mu_{i,3}-x_{30})}{\frac{\hat{\mathbf{s}}_{i,3}^2\hat{\mathbf{s}}_{i,2}r_1^2}{\hat{\mathbf{s}}_{i,1}^2}+\frac{\hat{\mathbf{s}}_{i,3}^2r_2^2}{\hat{\mathbf{s}}_{i,2}}+\hat{\mathbf{s}}_{i,2}r_3^2}\lesssim1,$$
	$$\frac{r_3^2(\mu_{i,2}-x_{20})}{\frac{\hat{\mathbf{s}}_{i,3}^2\hat{\mathbf{s}}_{i,2}r_1^2}{\hat{\mathbf{s}}_{i,1}^2}+\frac{\hat{\mathbf{s}}_{i,3}^2r_2^2}{\hat{\mathbf{s}}_{i,2}}+\hat{\mathbf{s}}_{i,2}r_3^2}\lesssim1.$$
	
	So we have $(splat_i)_{s_{i,2}}\lesssim{1}$, and when $r_2=r_3=0$, $(splat_i)_{s_{i,2}}\sim{1}$. According to the definition of equivalence, $(splat_i)_{s_{i,2}}\sim{1}$. And the same as other value of $j$.
	
	So we can find in certain $i$ and $k$, we have the realation that:
	$${\hat{\mathbf{s}}_{i,j}}\frac{{\partial}(T_k\hat{\mathbf{c}}_k\alpha_k)}{{\partial}\mu_{i,j}}\sim\frac{{\partial}(T_k\hat{\mathbf{c}}_k\alpha_k)}{{\partial}c_i}\sim\frac{{\partial}(T_k\hat{\mathbf{c}}_k\alpha_k)}{{\partial}o_i}\sim\frac{{\partial}(T_k\hat{\mathbf{c}}_k\alpha_k)}{{\partial}s_{i,j}}.$$
	Specially, the rotation attribute ${q_i^t}$ which is a quaternion array is updating as:
	$${q_i^{t+1}}={q_i^t}+\triangle{q_i^t},$$
	$$\lVert\triangle{q_i^t}\rVert=\lVert\frac{\frac{\partial{L}}{\partial{q_i}}+q_i^t}{\lVert\frac{\partial{L}}{\partial{q_i}}+q_i^t\rVert}-{q_i^t}\rVert\leq2\max(\lVert{q_i}\rVert)$$
	By the definition of a quaternionic array we have $\lVert{q_i}\rVert\leq1$, then we obtain $\triangle{q_i^t}\sim1$. So we can get
	$${\hat{\mathbf{s}}_{i}}\frac{{\partial}L}{{\partial}\boldsymbol{\mu_{i}}}\sim\frac{{\partial}L}{{\partial}c_i}\sim\frac{{\partial}L}{{\partial}o_i}\sim\frac{{\partial}L}{{\partial}s_i}\sim\triangle{q_i^t},$$
	so if we define the direction vector of $\triangle{q_i^t}$ as $r_{q,i}^t$, by the definition of partial derivatives, the updating of rotation attribute have
	$$\triangle{q_i^t}=\frac{{\partial}L}{{\partial}(q_i^t{\cdot}r_{q,i}^t)}.$$
	When the scales of Gaussians are small, we can get 
	$$\frac{{\partial}L}{{\partial}\boldsymbol{\mu_{i}}}\gg\frac{{\partial}L}{{\partial}c_i}\sim\frac{{\partial}L}{{\partial}o_i}\sim\frac{{\partial}L}{{\partial}s_i}\sim\frac{{\partial}L}{{\partial}(q_i^t{\cdot}r_{q,i}^t)}.$$
	
	That means the Gaussians will more willing to change their places to reduce the energy, which will more likely cause the large-scale random drift and leading the local minimum. 
	To achieve optimal results, we aim for all variables to change in a relatively consistent manner. To this end, it is natural to consider decelerating the changes in the positional attributes of the 3D Gaussians. Specifically, we formulate the 3DGS optimization procedure as the discretization of a Partial
	Differential Equation (PDE) and employ the viscosity coefficient, allowing spatial positions to absorb and gradually release the positional gradients of the 3D Gaussians.
	
	And due to the Gaussian function property, we have
	$$\sum_{splat_k\geq{\epsilon}}splat_k\sim1,$$
	where $\epsilon$ is the 0.99 confidence bound for the Gaussian function. So for the same 3D Gaussian $g_k$ at different scales $\hat{\mathbf{s}}_k$, we have
	$$\sum_{splat_k\geq{\epsilon}}\frac{{\partial}(T_k{\cdot}\hat{\mathbf{c}}_k{\cdot}\alpha_k)}{{\partial}\mu_{k}}=\sum_{splat_k\geq{\epsilon}}O(\frac{1}{\hat{\mathbf{s}}_k})T_k\hat{\mathbf{c}}_k\operatorname{Sig}(o_k)splat_k=O(\frac{1}{\hat{\mathbf{s}}_k}),$$
	and the position gradient $\frac{{\partial}L}{{\partial}\mu_{i,j}}$ of $g_k$ is proportional to $\sum\frac{{\partial}(T_k{\cdot}\hat{\mathbf{c}}_k{\cdot}\alpha_k)}{{\partial}\mu_{k}}$, so we have the relationship of position gradients between diffierent 3D Gaussians:
	$${\hat{\mathbf{s}}_{i,j}}\frac{{\partial}L}{{\partial}\mu_{i,j}}\sim{\hat{\mathbf{s}}_{k,j}}\frac{{\partial}L}{{\partial}\mu_{k,j}}.$$
	
	{\bfseries Observation.} 3DGS represents a complex scene as a set of 3D Gaussians. However, various 3DGS methods \cite{RN56, RN60} suffer from the common limitation of blurring and floaters due to the reconstruction of redundant and ambiguous geometric structures, leading to degraded rendering and reconstruction quality. We attribute the blurring and floaters to the occlusion of redundant large Gaussians and the ambiguity of small Gaussians, as shown in Fig. \ref{fig2}. The large 3D Gaussians fail to capture high-frequency details and tend to obstruct other Gaussians, resulting in redundancy and manifesting as blurring in the novel view. For small 3D Gaussians, due to the unstable gradient, floaters tend to appear in regions of the scene that are poorly observed, as the Gaussians tend to shift their positions toward observed views during the 3DGS optimization process, thereby resulting in ambiguous geometric structures.
	
	\subsection{The The velocity Voxel in Space}
	
	{\bfseries Building. }We aim to construct a loss function that considers the positional gradient field for a particle located at spatial position $\boldsymbol{\mu}$:
	$$v(\boldsymbol{\mu})=\sigma\frac{{\partial}L}{{\partial}\boldsymbol{\mu}}.$$
	
	However, since the attributes of the particles are unknown, this term cannot be directly calculated. Moreover, as the motion equations in 3DGS are based on the gradients of the scene rendering results, particles with different color attributes will exhibit different movement tendencies, typically lacking a linear relationship. Therefore, simply averaging attributes of the particles near a spatial location and then using this averaged set of attributes to compute the positional gradient is meaningless.
	
	We aim for the velocity field at $\boldsymbol{\mu}$ to indicate the most likely displacement of a Gaussian sphere at this location. Therefore, we choose to construct and update the velocity field by the local average velocity of $\boldsymbol{\mu}$, as shown in Eq.31. 
	
	This approach is mathematically meaningful: if we consider the positional gradients of 3D Gaussians as points in a three-dimensional space, the positional gradients of 3D Gaussians near $\boldsymbol{\mu}$ form a point cloud in this velocity feild. We want $v(\boldsymbol{\mu})$ to be positioned at the center of the largest cluster within this point cloud, which the arithmetic mean can achieve. Additionally, the arithmetic mean can counterbalance the impact of large-scale Brownian motion on the spatial velocity field caused by abrupt changes in positional gradients.Then we obtain the spatial velocity field $v(\boldsymbol{\mu})$.
	
	{\bfseries Total Impact of Gradient Field. }We renew the field by $\triangle\boldsymbol{v}^t_n=\frac{1}{{\mid}R_n^t\mid}\sum_{g_i{\in}R_n^t}{\triangle}\boldsymbol{\mu}_i^t$ in Eq.31. So we have the total impact of $\triangle\boldsymbol{v}^t_n$ in the field by adding it in every steps:
	$$I(\triangle\boldsymbol{v}^t_n)=\sum_{l\ge{t}}\triangle\boldsymbol{v}_n^t(l),$$
	where $\triangle\boldsymbol{v}_n^t(l)$ means portion of $\boldsymbol{v}_n^t(l)$ occupied by $\triangle\boldsymbol{v}^t_n$. So we have
	$$I(\triangle\boldsymbol{v}^t_n)=(1-\lambda_g\triangle\boldsymbol{v}^t_n\sum_{l\ge{t}}\lambda_g^{(l-t+1)}\rightarrow\triangle\boldsymbol{v}^t_n.$$
	
	Therefore, regardless of the coefficient $\lambda_g$, each updated vector will have a weight of 1 in the overall influence on the field throughout spacetime. At the same moment, the total weight of this vector on the gradient is always 1. Thus, no matter the chosen weighting coefficient, the value of this velocity field can naturally represent the magnitude of the gradient.
\end{document}